\documentclass{aa}  

\usepackage{graphicx}
\usepackage{float}
\usepackage{amsmath} % pour les environnements de matrices

\usepackage{txfonts}
\usepackage[colorlinks=true,allcolors=blue]{hyperref}
\usepackage{xcolor}

\begin{document}

   \title{Galaxy cluster count cosmology with simulation-based inference}

   \author{M. Regamey\inst{1} \and
          D. Eckert\inst{1} \and
          R. Seppi\inst{1} \and
          W. Hartley\inst{1} \and
          K. Umetsu\inst{2} \and
          S. Tam \inst{3} \and
          D. Gerolymatou\inst{1}
          }
   \institute{Department of Astronomy, University of Geneva, ch. d'Ecogia 16, 1290 Versoix, Switzerland \and
              Academia Sinica Institute of Astronomy and Astrophysics (ASIAA), No. 1, Section 4, Roosevelt Road, Taipei 106216, Taiwan \and Institute of Physics, National Yang Ming Chiao Tung University, 1001 University Road, Hsinchu 30010, Taiwan}
              
% \abstract{}{}{}{}{}
% 5 {} token are mandatory

\abstract{The abundance and mass distribution of galaxy clusters is a sensitive probe of cosmological parameters, in particular through the sensitivity of the high-mass end of the halo mass function to $\Omega_m$ and $\sigma_8$. While galaxy cluster surveys have been used as cosmological probes for more than a decade, the accuracy of cluster count experiments is still hampered by systematic uncertainties, such as the relation between survey observables and halo mass, the accuracy of the halo mass function, and the implementation of the survey selection function. Here we show that these uncertainties can be alleviated by forward modeling the observed cluster population with simulation-based inference. We construct a simulation pipeline that predicts the distribution of observables from cosmological parameters and scaling relations, and then train a neural network to learn the mapping between the input parameters and the measured distributions. We focus on fiducial X-ray surveys with available flux, temperature, and redshift measurements, although the method can be easily adapted to any available observable quantity. We apply our method to mock samples extracted from the UNIT1i N-body simulation and demonstrate the accuracy of our approach. We then study the impact of several important systematic uncertainties on the recovered cosmological parameters. We show that sample variance and the choice of the halo mass function are subdominant sources of systematic uncertainty. Conversely, the absolute mass scale is the leading source of systematic error and must be calibrated at the $<10\%$ level to recover accurate values of $\Omega_m$ and $\sigma_8$. However, the quantity $S_8=\sigma_8(\Omega_m/0.3)^{0.3}$ appears to be much less sensitive to the accuracy of the mass calibration. We conclude that simulation-based inference is a promising avenue for future cosmological studies from galaxy cluster surveys such as \emph{eROSITA} and \emph{Euclid} as it allows to consider all the available observables in a straightforward manner.}


\titlerunning{Cosmological parameters with SBI} % Titre abrégé pour l'en-tête
\authorrunning{M. Regamey et al.} % Auteurs abrégés pour l'en-tête

   \maketitle

\section{Introduction}

The discovery of the Universe's accelerating expansion significantly altered our understanding of its content and evolution, pushing us to develop new models \citep{Riess1998, Perlmutter1999}. The widely accepted model is the so-called $\Lambda$CDM model, which incorporates dark energy ($\Lambda$) and cold dark matter (CDM). Despite its mathematical simplicity, the $\Lambda$CDM model has been exceptionally successful, providing an impressive description of a wide range of astrophysical and cosmological observations \citep{Perivolaropoulos2022}. 

However, recent measurements have revealed anomalies in the $\Lambda$CDM model \citep[see][and references therein]{Abdalla2022,DiValention2025}. The cosmological parameters derived from primary anisotropies of the cosmic microwave background (CMB) radiation should be consistent with observations of the large-scale structure around us. Yet, several recent studies have highlighted discrepancies between early-Universe and late-Universe probes when assuming a $\Lambda$CDM model. There is a tension of more than 5$\sigma$ for the Hubble constant $H_0$ \citep[e.g.][]{Wong2020,Riess2022}, and a tension of about 3$\sigma$ on S8, a combination of the present day matter density $\Omega_m$ and the clumpiness parameter $\sigma_8$, which represents the amplitude of fluctuations within a co-moving sphere of 8 Mpc/h in diameter \citep[e.g.][]{Planck13XX, Bocquet2019, KiDS2021, DES2022}. If these tensions are confirmed, they could challenge the $\Lambda$CDM model and our current understanding of the Universe. Various significant systematic sources affect the different methods, underscoring the importance of refining our survey modeling techniques to obtain precise measurements from current data and investigate these tensions.

Galaxy clusters are the largest gravitationally bound structures in the Universe, making them highly sensitive probes of cosmological parameters \citep{Haiman2001, Allen2011, Dodelson2016}. The volume of galaxy clusters is filled by the intracluster medium (ICM), a diffuse medium composed of hot gas emitting mainly in the X-ray domain.  For this reason, X-ray surveys allow us at the same time to perform a deep census of the galaxy cluster population and its evolution, and to determine the properties of the ICM, such as its luminosity and temperature \citep[see][for a review]{Clerc2023}.

The galaxy cluster count method relies on the strong dependence of the halo mass function (HMF) on cosmological parameters such as $\sigma_8$ and $\Omega_m$, \citep{Vikhlinin2009, Planck13XX, Mantz2015b, Bocquet2019, Garrel2022, Chiu2023}. The HMF quantifies the number density of collapsed halos per unit mass and volume $\frac{dn}{d\ln(M)}$ and expresses it as a function of their mass $M$. Specifically, $\sigma_8$ defines the high mass cutoff of the HMF whereas changing $\Omega_m$ impacts the normalization of the mass function, such that observations of galaxy clusters provide a valuable tool to constrain these key cosmological parameters. To construct the HMF from an observed cluster survey, determining their halo masses is a major challenge, as these masses are not directly observable \citep{Pratt2019,Grandis2021}. This difficulty in measuring halo masses represents a significant hurdle in the study of galaxy clusters. Numerous studies have estimated the scaling relations between the cluster mass and integrated ICM properties such as temperature and luminosity \citep{Maughan2007, Pratt2009, Mantz012010, Mantz082010, Allen2011, Maughan042012, Maughan072012, Mantz2016, Pratt2019}. Since the gas temperature is a direct probe of the potential well, the presence of hot gas is a clear indication that the underlying halo is massive. Similarly, the total X-ray luminosity depends on the total amount of gas in the system, which scales with halo mass \citep[e.g.][and references therein]{Lovisari2021}. The combination of several proxies such as ICM temperature and luminosity thus provides valuable information on the mass of the underlying halos \citep[e.g.][]{Giodini2013, Ettori2015, Lovisari2020}.

Recovering accurate cosmological parameters from galaxy cluster counts requires precise knowledge of survey properties, such as selection function or mass calibration. Some of the key steps in modeling the observed galaxy cluster population are difficult to treat analytically and thus necessitate approximations in fully analytic formations. For this reason, forward modeling the distributions of observables generally outperforms classical methods based on a direct reconstruction of the HMF \citep[e.g.][]{Clerc2012,Pierre2017,Valotti2018}.

Here, we propose a forward modeling method for the joint reconstruction of cosmological parameters and galaxy cluster scaling relations using simulation-based inference (SBI). Given a set of parameters describing the cosmological model and the mass-observable scaling relations, we generate mock cluster datasets and train a deep learning algorithm to map the relation between the parameters and the summary statistics of the survey \citep[see also][]{Tam2022, Kosiba2025, Zubeldia2025}. Our method is fully numerical, which allows us to model exactly important effects such as the survey selection function, which can usually not be expressed analytically \citep{Pacaud2006, Mantz2015b,Garrel2022}. By generating simulated samples with observable properties, we can apply the exact numerical selection function in the observable space in which it is defined.

We jointly model the redshift, flux and temperature distributions of the detected cluster sample, which helps break the degeneracy between $\Omega_m$ and $\sigma_8$. Moreover, our method is easily adaptable to any additional observables. These can be self-consistently modeled by introducing additional scaling relations, that can be either fitted or marginalized over. Another advantage of our method is that we simulate fluxes within a fixed aperture, which makes no assumption on the system's mass.

In this paper, we present our forward modeling pipeline and SBI method. We validate the pipeline using mock samples extracted from N-body simulations and test the impact of the dominant sources of systematic uncertainty on our results. The paper is organized as follows. In Sect. \ref{sec:Mock sample description} we describe the mock sample used to validate our method. Sect. \ref{sec:Forward modeling} details the different steps of our pipeline for generating cluster samples from input parameters. We then discuss the inference results in Sect. \ref{sec:Results}, including the posterior distributions, a goodness-of-fit analysis, and a coverage test. In Sect. \ref{sec:Discussion}, we compare our approach with traditional methods and explore the influence of systematic effects, before concluding in Sect. \ref{sec:Conclusion}.

\section{Mock sample description} \label{sec:Mock sample description}

As a validation sample for our methodology, we use the mock galaxy cluster catalog described in \cite{Seppi2022}. Their simulation was generated to predict the population observed by \emph{eROSITA}, and is therefore suitable as a benchmark for this work. In this section, we summarize the main properties of this simulation. It is based on a light cone generated from combining multiple snapshots of the UNIT1i dark matter only simulations \citep{Chuang2019}. These simulations assume a flat $\Lambda CDM$ cosmology \citep{PlanckCollaboration2016}, with fiducial parameters $H_0 = 67.74$ km s$^{-1}$ Mpc$^{-1}$, $\Omega_{\rm m0} = 0.3089$, $\Omega_{\rm b0} =
0.048206$, and $\sigma_8$ = 0.8147. The comoving box size of 1 $Gpc/h$ and the particle mass resolution of $1.2\times 10^9 M_\odot /h$ allow a very detailed description of the massive haloes in the simulation.

Shells of individual snapshots are combined into a light cone, and the comoving distance is converted into redshift also accounting for peculiar velocities. Galaxy clusters are painted onto dark matter haloes following \cite{Comparat2020}. The method relies on generating mock data from real observations. A collection of 326 galaxy clusters combining XMM-XXL \citep{Pierre2016}, HIFLUGCS \citep{Reiprich2002}, X-COP \citep{eckert2019}, and SPT-Chandra \citep{sanders2018} is used to generate a covariance matrix between emission measure profile, temperature, mass, and redshift. Extracting samples from the covariance matrix allows the generation of emission measure profiles and temperatures with the proper correlation with halo mass and redshift, as seen in real observations of galaxy clusters with high signal-to-noise ratio. Profiles and temperatures are assigned to dark matter haloes in the mock light cone by a nearest neighbor search in mass and redshift. Input values of X-ray luminosities are obtained by integrating the emission measure profiles accounting for the cooling function in the 0.5-2.0 keV band \citep[see][for more details]{Comparat2020}. Since the profiles are rescaled by $R_{\rm 500c}$, it is straightforward to obtain the luminosity within apertures of $R_{\rm 500c}$ and a fixed physical aperture of 300 kpc by changing the limits of integration. Finally, X-ray fluxes in the observer frame are derived from the luminosity distance given the cluster redshift and accounting for the K-correction as a function of redshift, temperature, and absorption properties obtained from neutral hydrogen maps from \cite{Hi4PI2016}.

Since the model is based on massive clusters, it naturally tends to over-predict the X-ray luminosities in the regime of galaxy groups for $M_{\rm 500c}<5\times 10^{13}$ $M_\odot$. \cite{Seppi2022} overcome this limitation by recalibrating the X-ray luminosity in the galaxy group regime based on the scaling relation with stellar mass from \cite{Anderson2015}. The authors showed that the correction is negligible for $M_{\rm 500c}> 10^{14}$ $M_\odot$. Therefore, any assumption about this correction does not affect this work, where we focus on galaxy clusters. Overall, the method accurately reproduces the cluster number density as a function of flux \citep{Finoguenov2020, Liu2022}, and scaling relations between X-ray cluster observables and mass \citep{Lovisari2015, Lovisari2020, Schellenberger2017, Bulbul2019, Adami2018}. The mock cluster catalog is publicly available\footnote{\url{http://cdsarc.u-strasbg.fr/viz-bin/cat/J/A+A/665/A78}}. For additional details we refer the reader to \citet{Comparat2020} and \citet{Seppi2022}.

\subsection{Mock Sample Definition}

\begin{table}

\centering
\caption{Mock sample selection function in terms of redshift and mass cuts.}
\begin{tabular}{c c}
\hline\hline
& \textbf{Selection function}\\
\hline
\textbf{redshift} & $0.1 < z < 0.6$ \\
\textbf{mass} & $M_{\text{500c}} > 10^{14.2}\ M_\odot$ \\
\textbf{N$_{\rm halo, TOT}$} & $1\,116\,758$ \\
\textbf{N$_{\rm halo, SEL}$} & $30\,939$ \\
\hline
\end{tabular}
\vspace{0.5em}

\tablefoot{We also report the total number of clusters in the half-sky light cone simulation and the ones filtered by the ideal selection.}
\label{tab:selfunc}
\end{table}

\begin{figure}
\centering
\includegraphics[width=\columnwidth]{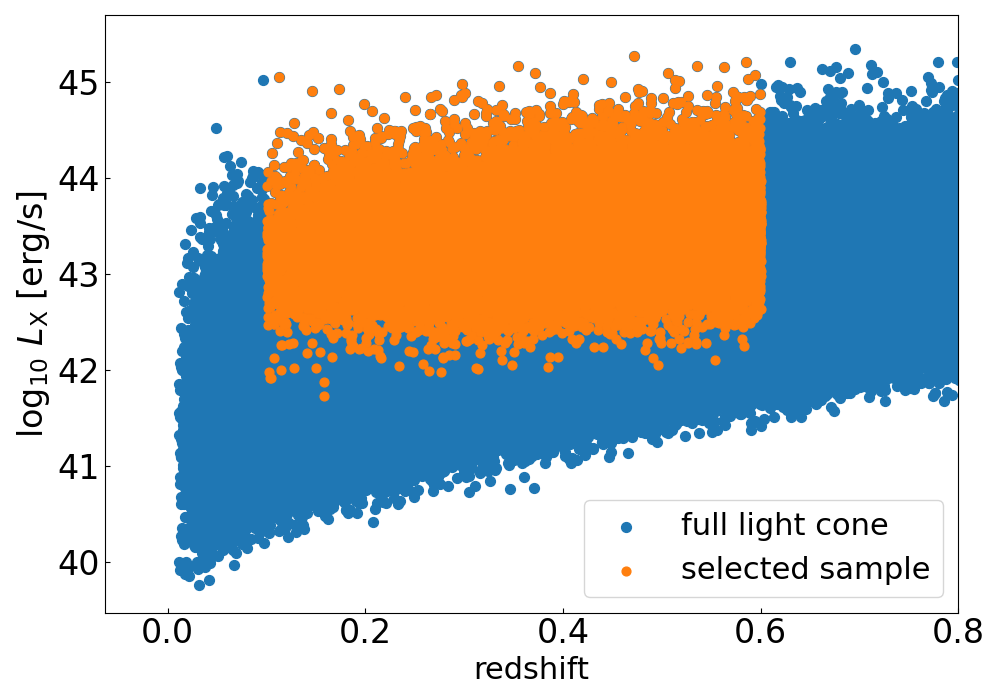}
\caption{Luminosity-redshift distribution for clusters in the simulation of \cite{Seppi2022}. The blue color refers to the full population, the orange to the one after the application of the ideal selection function.}
\label{Fig:Lx_z_mock}
\end{figure}

Cosmology experiments using cluster counts require an accurate estimate of the selection function. This is a key ingredient in the forward modeling of a population generated from a theoretical HMF to an observed sample. A biased estimate of the selection function will inevitably have repercussions on the cosmological constraints. To test our cosmological pipeline using the simulation, we introduce a simple selection function based on true halo properties rather than observables. This allows us to validate our pipeline without introducing possible systematics related to the selection function itself. However, we stress that the method developed in this paper is suitable for any selection function, as long as it is known at a sufficient level of accuracy.

Starting from the full simulated sample, we apply strict cuts in mass and redshift and mock an ideal selection function. We include all clusters more massive than $10^{14.2}\ M_\odot$ and between redshifts 0.1 and 0.6 (see also Table \ref{tab:selfunc}). After filtering, we obtain a sample of $30\,939$ clusters. Figure \ref{Fig:Lx_z_mock} shows the impact of the ideal selection on the distribution of clusters as a function of X-ray luminosity and redshift. The ideal mass selection makes our sample very close to a volume limited sample with a fixed luminosity cut. However, we notice that some clusters scatter around values of $10^{\rm 42}$ erg/s. This is due to the intrinsic scatter between X-ray luminosity and halo mass, which is naturally accounted for by our formalism, as explained in the next section. The impact of the selection on the mass and redshift distributions can be found in Appendix \ref{sec:Mock selection impact}.

\section{Forward modeling} \label{sec:Forward modeling}

In this section, we detail the steps of our simulation pipeline that enables us to generate galaxy cluster samples in terms of temperature, flux, and redshift distributions, based on input cosmology and the mass-luminosity scaling relation parameters. We assume a flat $\Lambda$CDM model whereby the evolution of the scale factor is uniquely determined by $\Omega_m$ and $H_0$. 

\begin{figure*}
\centering
\centerline{\resizebox{\hsize}{!}
{\includegraphics{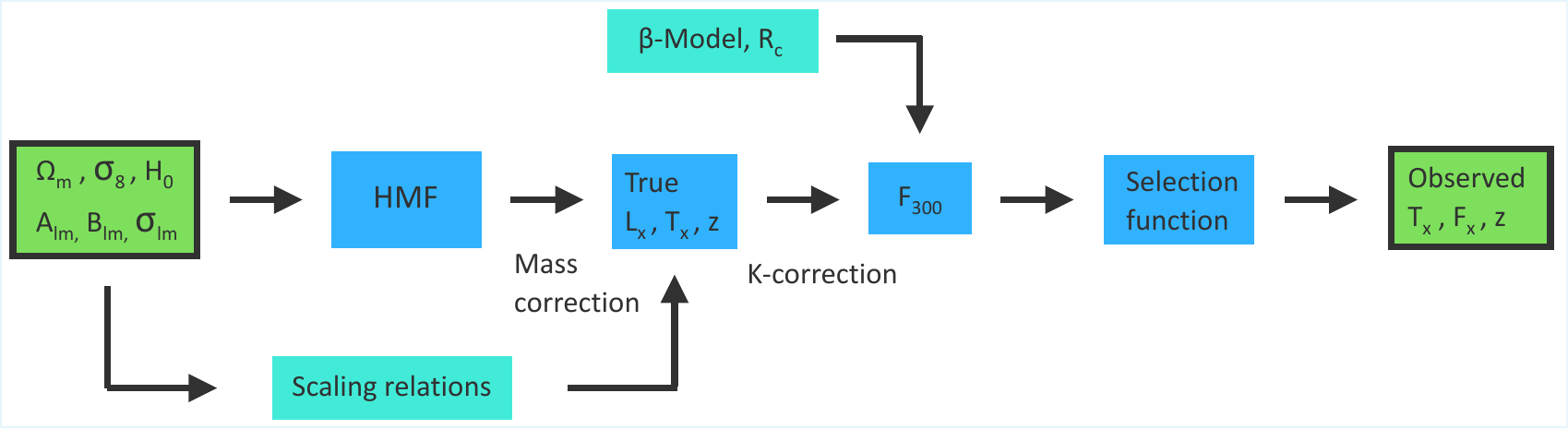}}}
\caption{Forward modeling pipeline used to generate simulated samples of galaxy clusters from input parameters. We begin with a set of input cosmological parameters: $\Omega_m$, $\sigma_8$, and $H_0$. From these parameters, we generate cluster masses within a fiducial volume slice of fixed redshift (see Section \ref{sec:Survey volume}) using the HMF (see Section \ref{sec:Mass function}). Before applying the scaling relations, we ensure that the cluster masses are defined consistently with the cosmology associated with these relations (see Section \ref{sec:Mass correction}). Luminosities and temperatures are then derived from these masses through scaling relations (see Section \ref{sec:Scaling relation}). We apply a $k$-correction to derive observer-frame luminosities (see Section \ref{sec:k-correction}) and use a $\beta$-model to rescale the predicted flux into a fixed physical aperture (see Section \ref{sec:Beta model and core radii}). Finally, we apply the selection function of the observed sample on the relevant observables to produce a mock catalog of galaxy clusters that is directly comparable with the observed sample (see Section \ref{sec:Selection function}).}
\label{Fig:pipeline}
\end{figure*}

The adopted modeling scheme is summarized in Fig.\ref{Fig:pipeline}. The pipeline incorporates our physical knowledge of cluster physics and the expected survey observables. Importantly, the approach described in this paper allows us to easily adapt our pipeline to the observables at hand and integrate additional physical insights as needed. For instance, our procedure can be easily expanded to additional observables such as weak lensing shear \citep{Tam2022} or richness.

\subsection{Survey volume} \label{sec:Survey volume}

We start by calculating the survey volume in our input cosmology within redshift bins spanning the redshift range covered by the survey of interest. For the mock sample (Sect. \ref{sec:Mock sample description}), we define redshift slices ranging from 0.1 to 0.6 in increments of 0.01. To ensure accurate modeling, the redshift bins are selected to be sufficiently fine, such that the mass function is assumed to be independent of redshift within each redshift slice. The binning strategy and the chosen redshift range can be easily adapted to any survey of interest. In this work, we consider two survey configurations with areas of 50 deg$^2$ similar to the XMM-XXL survey \citep{Pierre2016} and 1000 deg$^2$ similar to Subaru/HSC-SSP survey \citep{Aihara2018}, enabling comparisons with surveys of varying sky coverage. The volume of each slice is calculated as

\begin{equation} \label{Eq:Volume}
    V(z_{min}, z_{max}) = \frac{d\Omega}{4\pi}\int_{z_{min}}^{z_{max}}\frac{dV}{dz}(\Omega_m, H_0) \, dz
\end{equation}

where $dV/dz$ is the differential comoving volume, which depends on cosmology, and $d\Omega$ is the survey area in steradians \citep{Hogg1999}. We then generate clusters inside this fiducial volume by drawing objects from the mass function. 

\subsection{Mass function} \label{sec:Mass function}

Our redshift bins are chosen to be narrow enough such that the mass function is approximately independent of redshift within each slice. The mass function and it's evolution are calculated at the mean redshift of each slice, using the \texttt{colossus} Python package \cite{Diemer2018} and the \cite{Tinker2008} model. The \citet{Tinker2008} model was found to provide the closest representation of the HMF in the UNIT1i N-body simulation, which motivates our baseline choice. The effect of using a different mass function model, which would be less closely aligned with our mock data, is presented in Sect. \ref{sec:Mass function model comparison}. We generated halos within each redshift slice by drawing masses from the HMF within the mass range between $1.5\times10^{14}M_\odot$ and $10^{15.5}M_\odot$ with a step size of 0.01 dex. Detecting a halo beyond these limits within our survey volume is highly improbable, thus our choice allows us to sample the entire mass range of interest. Masses are generated within an overdensity $\Delta_c = 500$. This choice arises from the definition of mass in the mass-temperature scaling relation (M-T), see \ref{sec:Scaling relation}, which is the typical overdensity within which weak lensing (WL) masses are calculated. We define overdensity masses $M_\Delta$ and radii $R_\Delta$ as the mass and radius within which the mean density is $\Delta_c$ times the critical density of the Universe,

\begin{equation} \label{Eq:rho}
  \overline{\rho} = \frac{M_{\Delta}}{4/3\pi R_{\Delta}^3}=\Delta_c \rho_c(z)
\end{equation}

with $\rho_{c}(z)$ the critical density of the Universe at the redshift of the system. The total number of halos within each redshift slice is given by the integral of the mass function over the mass and the survey volume,

\begin{equation} \label{Eq:nbr halo}
\begin{split}
  N_{halo}(z_{min}, z_{max}) &= \frac{d\Omega}{4\pi}\int_{z_{min}}^{z_{max}}\frac{dV}{dz}(\Omega_m, H_0) \, dz\int_{M_{min}}^{\infty}\frac{dN}{dMdV}(z)\, dM\\ &\approx  V(z_{min},z_{max})\int_{M_{min}}^{\infty}\frac{dN}{dMdV}(\langle z \rangle)\, dM
\end{split}
\end{equation}

The approximation made in this equations follows from our assumption that the mass function does not evolve within the small redshift interval of any of our given bins.

The generated number of clusters within each redshift slice is given by a Poisson realization of the expectation value in Eq.\ref{Eq:nbr halo}. While the redshift of each slice is considered fixed for the purpose of mass generation using the HMF, the actual sources are distributed across a continuous range of redshifts. This is done by sampling from a distribution weighted by the differential comoving volume within each bin, allowing for a more realistic spread of source redshifts rather than assigning a single redshift to all sources in the slice.

\subsection{Mass correction} \label{sec:Mass correction}

The masses generated from the mass function are given within an overdensity $\Delta_c = 500$. Since $\rho_c(z) = \rho_{c,0} E(z)$ is cosmology dependent, with $E(z)=H(z)/H_0$ the expansion factor, the definition of $M_{\Delta}$ is also cosmology dependent. To consistently apply scaling relations for galaxy clusters (see \ref{sec:Scaling relation}) and generate observables, we must first convert our simulated masses into the cosmology assumed for these relations. 

In our case, the mock catalog, and therefore the mass observable we use, was extracted from a simulation that assumed a Planck 2015 cosmology \citep{PlanckCollaboration2016}. In the general case, we assume that constraints on the M-T scaling relation are available in a fixed cosmology of choice, and within our pipeline we wish to convert the generated masses into the cosmology that was assumed to estimate the scaling relation. 

Most recent estimates of the M-T scaling relation employ WL as a mass calibration method \citep{Mantz2016, Mulroy2019, Sereno2020}, as the lensing signal is independent of a cluster's dynamical state \citep{Umetsu2020b}. Therefore, we assume that the relation between WL mass and gas temperature has been previously calibrated in a fixed cosmology and we use the background galaxy distribution to convert the simulated masses to the corresponding cosmology. We convert the mass calibration from one cosmology to another by taking into account the impact of the change in redshift - distance relation of the source lensed galaxies. Here, we assume that our survey has properties similar to Subaru/HSC-SSP, leading to a redshift distribution that is similar to the one obtained in \cite{Umetsu2020a}.

In general, WL mass calibration employs either photo-z cuts or color–color cuts to identify foreground cluster and background galaxies \citep{Medezinski2018}. In this work, we adopt the photo-z cut approach of \cite{Umetsu2020a}, in which the photometric redshift probability distribution function (PDF), $P(z)$, of each galaxy is used to select background sources behind a given cluster. Assuming a survey quality similar to HSC-SSP, we can estimate a realistic distribution of background galaxies as a function of cluster redshift. We assume that the redshift distribution of background galaxies follows a known distribution $P(z)$. The cosmological dependence of the WL mass of the cluster inferred from these data can be described using the analytic formula introduced by \citet{Sereno2015},

\begin{equation} \label{Eq:Mass conv}
  M_{500c, WL} \propto \left(\frac{D_{ds}}{D_s}\right)^{-3/2} \cdot H(z)^{-1}
  \end{equation}
  
where $D_s$ and $D_{ds}$ represent the source and lens-source angular diameter distances, respectively. $H(z)$ is the Hubble function evaluated at the redshift $z$ of the cluster. Since the distance ratio between $D_s$ and $D_{ds}$ varies for each background source, we compute the average of Eq.\ref{Eq:Mass conv} over all background sources weighted by the number of background sources at each redshift, thus incorporating the stacked photo-z PDF $\langle P\rangle(z)$ of the selected background galaxies behind the cluster:

\begin{equation} \label{Eq:Mass conv2}
\left \langle  \left(\frac{D_{ds}}{D_s}\right)^{-3/2}\right\rangle\cdot H(z)^{-1} = \frac{\sum \langle P\rangle(z) \cdot \left(\frac{D_{ds}}{D_s}\right)^{-3/2}}{\sum \langle P\rangle(z)}\cdot H(z)^{-1}
  \end{equation}
  
For each specific simulated cluster redshift in our pipeline, we search for the cluster studied by \citet{Umetsu2020a} with the closest redshift and associate the $\langle P\rangle(z)$ distribution of the nearest neighbor to that cluster. Then, after calculating Eq.\ref{Eq:Mass conv2} for both our simulated cosmology and the cosmology within which the scaling relation was estimated, we obtain a mass correction factor for the target cluster which is simply the ratio of these two quantities. This factor accounts for the cosmology-dependent change in the weak lensing mass estimate, allowing us to transform the simulated mass into the equivalent lensing mass in the target cosmology. Assuming that the $\langle P\rangle(z)$ distribution depends only on the redshift of our cluster, we compute the correction factor across a redshift grid ranging from 0.1 to 0.7 with a step of 0.05, and then interpolate for each generated redshift. We thus make the assumption that the depth of observations for WL is uniform across the survey, which is reasonable given the high homogeneity of the HSC-SSP survey depth \citep[$i \sim 24.5$ ABmag,][]{Aihara2018}. Finally, using the obtained correction factor, we convert our generated masses into the cosmology that was assumed for the estimation of the M-T relation, which we then use to generate mock observables.

\subsection{Scaling relations} \label{sec:Scaling relation}

\textbf{The mass-temperature scaling relation}
In the self-similar model \citep{Kaiser1986} and its extensions \citep{Ettori2015, Ettori2020}, ICM properties are expected to be tightly correlated with halo mass. In particular, the temperature is a tight proxy for the halo mass because of its dependency on the depth of the potential well \citep[e.g.][]{Nagai2007,Truong2018,Pop2022,Braspenning2024}. In this work, we assume that the M-T scaling relation is known, with an uncertainty and a scatter that we propagate into our simulation pipeline. In a more general case, our method can be adapted to the situation where WL studies are directly available, by fitting jointly the scaling relations and the cosmology.

Our mock sample was extracted from N-body simulations considering dark matter only, such that the gas properties had to be painted on using a semi-analytic approach (see Sect.\ref{sec:Mock sample description}). The semi-analytic model follows \cite{Comparat2020} to generate cluster properties from their mass, such as their temperature $T_{500}$ and luminosity $L_{500}$. We fit the M-T relation of the clusters in the mock catalog with a power law and found that the relation can be well described by

\begin{align} \label{Eq:MassTx}
  T_{500} &= 2.104 \cdot \left( \frac{M_{500}}{10^{14} M_{\odot}} \right)^{0.576} \cdot E(z)^{0.458} \text{keV}
\end{align}

with $M_{500}$ and $T_{500}$ the mass and the temperature in an overdensity of 500, respectively. We therefore implement this relation in our pipeline to generate temperatures from the simulated masses. Finally, the simulated temperatures are obtained as a Gaussian realization around the mean value, with a log-normal scatter of 0.07 dex that matches the scatter implemented in the \cite{Comparat2020} model. Both theoretical and observational studies have shown that this level of scatter is typical, as demonstrated e.g. by \citet{Truong2018}.

\par
\vspace{5mm}
\textbf{The mass-luminosity scaling relation}
The luminosity of a halo is linked to its gas fraction, and therefore to its mass \citep[e.g.][]{Pratt2009}. We model the mass-luminosity (M-L) scaling relation as a power law with a log-normal scatter,

\begin{equation} \label{Eq:ML} L_{500} = E(z)^{7/3} \cdot \left(\frac{M_{500}}{10^{14}M_\odot}\right)^{A_{lm}} \cdot B_{lm}. \end{equation}

Here $A_{lm}$ and $B_{lm}$ are the slope and the normalization of the relation, respectively. $L_{500}$ represents the luminosity in the 0.5-2 keV energy band within an overdensity of 500. The $7/3$ factor on $E(z)$ arises from the definition of $L_{500}$, which is the cylindrical luminosity integrated along the line of sight within a projected radius $R_{500}$; see \citet{Lovisari2022} for more details.

While in principle the slope and normalization of the scaling relation can be predicted from the self-similar model, several studies found the M-L relation to be steeper than the self-similar expectation \citep{Pacaud2007, Pratt2009, Bulbul2019, Lovisari2020}. However, the relation can still be well described by a power law, as there is no evidence for a break in the relation down to group scales \citep{Anderson2015,Zhang2024,Wood2025}. Therefore, we treat the slope $A_{lm}$, the normalization $B_{lm}$, and the scatter $\sigma_{lm}$ of the power law as free parameters in our pipeline to account for deviations from the self-similar model. As a result, the cosmological constraints we obtain will be marginalized over the uncertainty in the M-L relation, and the relation itself will be determined simultaneously with the cosmological parameters.

\subsection{K-correction} \label{sec:k-correction}

The mass-luminosity relation defined in Eq.\ref{Eq:ML} is defined in the 0.5-2 keV band in the rest frame of each cluster, whereas the fluxes extracted from the survey are estimated in the corresponding band in the observer frame. Thus, a correction from rest frame to observer frame must be applied for every simulated system, given its input redshift and temperature.

Going from rest frame to observer frame, the spectrum is shifted to lower energies due to the expansion of space between the source and the observer. Given the temperature of a simulated object, we use the Astrophysical Plasma Emission Code \citep[APEC,][]{Smith2001} model to calculate the flux ratio between the observed band and the band in the rest frame. We compute the flux in the observer frame accounting for redshift in the energy limits of the integral: 
\begin{equation}
F_{\text{obs}} = \int_{E_{\text{min}}(1+z)}^{E_{\text{max}}(1+z)} F_E \, dE.
\end{equation}
Since this conversion does not depend on cosmology, we can define the quantity $K$ as the ratio of the observed flux to the rest frame one, which is equivalent to the ratio of luminosities in observer frame and rest frame. We assume a constant gas metallicity of $0.3Z_\odot$ with respect to the \citet{Asplund:2009} Solar abundance table and a fixed absorption column density $N_H=10^{20}$ cm$^2$, although in a general case the K-correction can be easily calculated as a function of $N_H$. Given these assumptions, the conversion factor depends only on redshift and temperature. We calculate the conversion factor over a grid of redshifts and temperatures, and interpolate over the grid to obtain the $K$ factor of each generated cluster. The observed flux $F_{x}$ is then obtained from the rest-frame luminosity as

 \begin{equation} \label{Eq:F500}
  F_{x,500} = \frac{L_{500} }{4 \pi d_L^2} \cdot K(T,z)
  \end{equation}
  
with $K(T,z)$ the K-correction at the temperature and redshift of each source, and $d_L$ the luminosity distance.
\subsection{Beta model and core radii} \label{sec:Beta model and core radii}

The luminosity we obtain from the M-L scaling relation is integrated within $R_{500}$, which depends on the mass of the cluster. Given a model for the brightness distribution of the source, we can convert $F_{x,500}$ into the flux estimated within a fixed physical aperture that is directly comparable to the measured fluxes. For a surface brightness distribution $I(r)$, the flux integrated within a radius $R$ becomes

\begin{equation} \label{Eq: Fluxinrad}
F(<R) = \int_0^R 2\pi rI(r) \, dr
\end{equation}

We model the brightness distribution of the simulated clusters using the beta model \citep{Cavaliere1976}, which describes the expected emissivity distribution of an isothermal sphere in hydrostatic equilibrium:

\begin{equation} \label{Eq: betamodel}
I(R) = I_0 \left(1+\left(\frac{R}{R_c}\right)^2\right)^{-3\beta+1/2} 
\end{equation}

The shape of the distribution is governed by a single parameter $R_c$, that is, the radius around which the profile transitions from a flat core to a power-law decrease. While clusters generally deviate from isothermal spheres, the beta model is known to provide a good approximation of the brightness distribution, with deviations at the level of $\sim10\%$ \citep{Käfer2019}. Given values of $R_{500}$ and $F_{x,500}$, we can obtain the normalization of the beta model, $I_0$, and thus a flux within any aperture. Studies of the brightness distribution in galaxy clusters showed that the $\beta$ parameter is usually around 2/3 \citep[e.g.][]{Mohr1999, Chen2007, Eckert2011, Käfer2019}. Assuming $\beta=2/3$, the flux enclosed within radius $R$ thus becomes

\begin{equation} \label{Eq: analytint}
F(<R) = 2\pi I_0 R_c^2\left[1-\left(1+ \left(R/R_c\right)^2\right)^{-1/2} \right]
\end{equation}

To account for realistic structural variations among the cluster population, we simulate the core radius based on the distribution of values observed in well-studied clusters, see Appendix \ref{sec:core radius distribution} for a full description of the procedure.

\cite{Käfer2019} demonstrated that the luminosities and core radii of clusters are anti-correlated, which can be explained by the dependence of the surface brightness profile on the dynamical state \citep[e.g.][]{Leccardi2010,Rossetti2017,Andrade2017}. We inject this knowledge into our pipeline by generating the scatter in the luminosities and the core radii simultaneously from a multivariate Gaussian distribution, assuming a correlation coefficient of $-0.43$ between $R_c$ and $L_{500}$ \citep{Käfer2019}.

In practice, the mass of each system is not known a priori and the fluxes within $R_{500}$ cannot be easily estimated observationally. Therefore, we assume that the fluxes have been extracted within fixed physical apertures. The choice of the aperture is arbitrary, but fixed and known in advance. Here, following \citet{Giles2016} we assume an aperture of 300 kpc, which should encompass the majority of the flux for the objects we detect. We stress that our method can work for any choice of aperture as long as the aperture is treated consistently, meaning that we can adapt our simulated aperture to any observed survey. From the luminosity, the core radius and the beta model, we can easily and analytically calculate a flux within any aperture and inject it into the simulation pipeline to replicate a fixed aperture in a fixed cosmology.

\subsection{Selection function} \label{sec:Selection function}

To ensure that our simulations are comparable to the data, the selection function of the survey needs to be implemented. With our approach, we can generate observables as they are used for the selection function of a survey. Our methodology enables us to generate these parameters directly, ensuring that the selection function is accurately applied and that the sample we produce closely mirrors what would be observed by the survey. In the case of the mock sample used in this study and described in \ref{sec:Mock sample description}, the selection function is simply a cut in mass, $M_{\text{500c}} > 10^{14.2}\ M_\odot$, and redshift $0.1 < z < 0.6$. However, we stress that the approach proposed here allows us to implement any selection function in an exact way, provided that the selection is sufficiently well understood.

\subsection{Optimization with SBI} \label{sec:Optimization with SBI}

With the forward modeling pipeline in hand, an optimization scheme is required to compare the simulated samples with the observational dataset and determine the best-fit parameters. We use the \texttt{sbi} Python package \citep{Tejero_Cantero2020}, which implements the Density Estimation Likelihood-Free Inference (DELFI) algorithm. The code generates a large number of simulations, spanning a pre-defined range of parameter values, allowing us to vary different input parameters of a specifically defined model. We then use the sequential neural posterior estimation \citep[SNPE,][]{SNPE} method  to learn the mapping between the input parameters and the simulated samples and return posterior distributions of each parameter. Once the model is trained, we apply it to the observed dataset and sample the parameters directly from the learned posterior distribution, represented by a neural density estimator. We apply flat priors on the parameters of interest (see Table \ref{tab:PriorsSBI}) and generate 10,000 simulations using parameters drawn from the prior distribution.

\begin{table}
\centering
\caption{Priors used for each input parameter of the pipeline in the SBI optimization algorithm: the 3 cosmological parameters $\Omega_m$, $\sigma_8$ and $H_0$, and the 3 parameters from the M-L scaling relation $A_{lm}$, $B_{lm}$ and $\sigma_{lm}$ corresponding to the slope, the normalization, and the log-normal intrinsic scatter, respectively. $\mathcal{U}([min,max])$ represents a flat distribution.}
\begin{tabular}{c c c}
\hline\hline
Parameter & Prior\\
\hline
$\Omega_m$ & $\mathcal{U}([0.155, 0.5])$ \\
$\sigma_8$ & $\mathcal{U}([0.6, 1.2])$ \\
$H_0$ & $\mathcal{U}([65., 75.])$ \\
$A_{lm}$ & $\mathcal{U}([0.5, 3.])$ \\
$B_{lm}$ & $\mathcal{U}([-1., 1.])$ \\
$\sigma_{lm}$ & $\mathcal{U}([0., 2.])$ \\
\hline
\end{tabular}
\vspace{0.5em}
\label{tab:PriorsSBI}
\end{table}

\begin{figure}
\centering
\includegraphics[width=\columnwidth]
{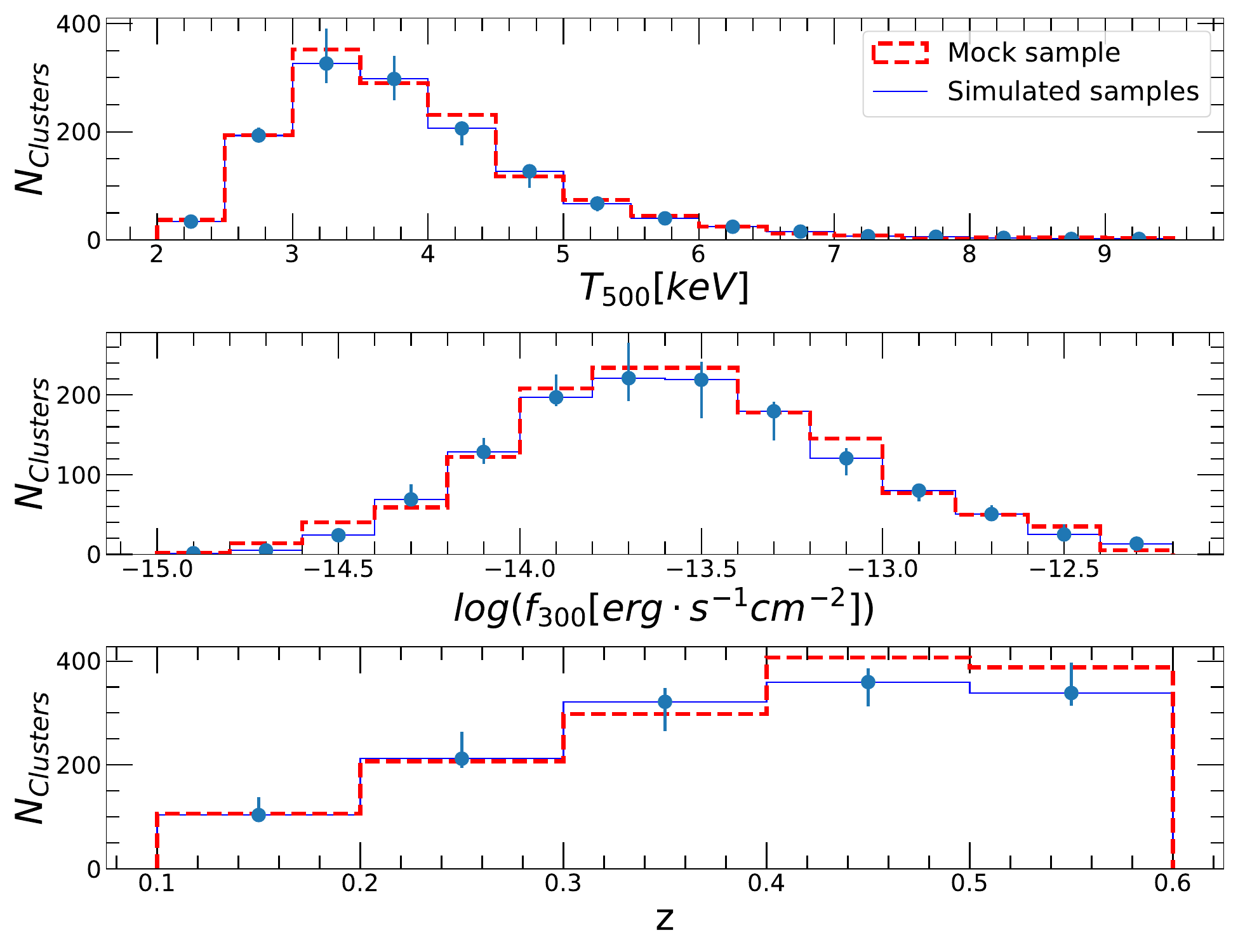}
\caption{Temperature, flux, and redshift histograms for 1000 deg$^2$ with the UNIT1i mock sample in red and results computed from the obtained posterior parameters in blue. The blue points show the median value for each bin across a range of simulated samples generated from the posterior distributions using our method with flat priors for each parameter and a uniform proposal distribution for training the model. The error bars show the standard deviation of the simulated samples in each bin.}
\label{Fig:errorbar_flatprop_1000}
\end{figure}

The \texttt{sbi} Python package requires both the observed dataset and the outputs of the simulations to have a fixed size. Since the number of generated clusters varies in each simulation, we compute fixed-size histograms of the temperature, flux, and redshift distributions using predefined binning, see Table \ref{tab:BinsSBI} and Fig.\ref{Fig:errorbar_flatprop_1000}. We then build a data vector made of the combination of all the histograms.
 
We tested several different configurations for the definition of the data vector, including a three-dimensional histogram of flux, temperature, and redshift, and a 2D histogram for the flux and temperature with a separate 1D histogram for the redshift distribution. We found that the 2D solution is more stable, thus for the remainder of the paper we adopt this configuration. For more details on the different setups, we refer to Appendix \ref{sec:Sbi configurations}.

\begin{table}
\centering
\caption{Bins used to create output temperature, flux and redshift histograms in the SBI optimization algorithm.}
\begin{tabular}{c c c}
\hline\hline
Parameter & Bins(min, max, step)\\
\hline
Temperature & (2, 10, 0.5) \\
Flux & (-15, -12, 0.2)\\
Redshift & (0.1, 0.7, 0.1)\\
\hline
\end{tabular}
\vspace{0.5em}
\label{tab:BinsSBI}
\end{table}

To train the model in SBI, the parameter space is explored by randomly sampling values from the prior distribution. In this case, we use the priors outlined in Table \ref{tab:PriorsSBI}, which specify plausible ranges for each of the parameters of interest, meaning the parameter space is explored uniformly across the six-dimensional cube set by the individual prior ranges. Uniform sampling ensures a broad exploration of all possible solutions, without biasing the search toward any specific region. 

To improve efficiency and reduce time spent exploring uninformative regions of the parameter space, we tried to enhance the sampling strategy by introducing a non-uniform proposal distribution. To do so, we first train a model on a sample generated from a uniform proposal and infer a posterior distribution. From the resulting posterior, we draw 100,000 samples, which are then used to compute a Fisher matrix. This matrix defines a multivariate normal distribution that serves as our updated proposal. Additional configuration details and results are provided in Appendix \ref{sec:Fisher proposal}.

\section{Results} \label{sec:Results}

In this section, we present the results of our tests evaluating the performance of our method and the SBI approach. Since the cosmological parameters used to generate the UNIT1i mock sample are known (see Table \ref{tab:parameter results}), we can rigorously assess the accuracy of our method in recovering these parameters.

\subsection{Validation of SBI-Inferred Parameters} \label{sec:Validation of SBI-Inferred Parameters}
We test the performance of the SBI method applied to our pipeline through its ability to reproduce the observed distributions of temperature, flux, and redshift. To achieve this, we selected a region of 1000 deg$^2$ from the half-sky light cone and attempted to recover the input parameters using the model trained with uniform proposals on each parameter corresponding to the chosen prior ranges (see Table \ref{tab:PriorsSBI}). To verify that the model provides an adequate representation of the data, we then randomly selected 1000 parameter sets from the posterior distribution and generated a simulated sample for each using our pipeline. The median and dispersion of the resulting samples were then computed in each bins of redshift, temperature, and flux. Fig.\ref{Fig:errorbar_flatprop_1000} shows how these distributions compare with those of the mock sample.

This comparison clearly demonstrates that, with flat priors and a uniform proposal over 1000 deg$^2$, our method effectively converges to parameters that replicate the temperature, flux, and redshift distributions extracted from the mock sample. However, small deviations between the simulated samples and the data can be observed, particularly at redshifts $z>0.4$; the following section investigates the quality of the fit to determine whether these discrepancies are consistent with statistical fluctuations or indicative of a systematic effect.

\subsection{Goodness-of-fit} \label{sec:Goodness-of-fit}

\begin{figure}
\centering
\includegraphics[width=\columnwidth]
{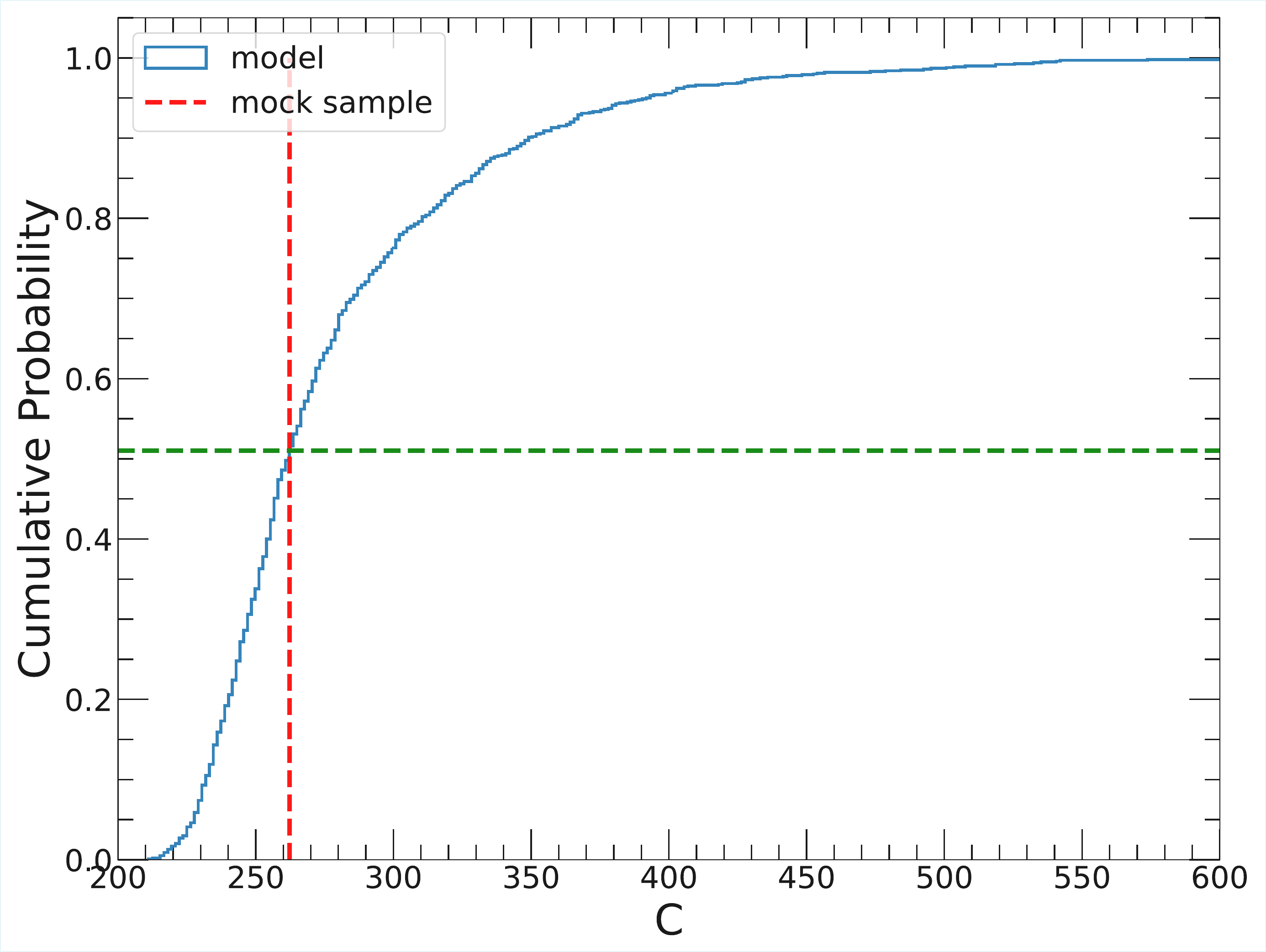}
\caption{Cumulative distribution of the test statistic $C$ obtained from 1000 realizations coming from random points in the parameter posterior distributions for the 1000 deg$^2$ test. The red dashed line is the $C$ value of the mock sample.}
\label{Fig:Gof}
\end{figure}

A key question in our analysis is whether the observed data are compatible with a random realization of our simulation pipeline from the posterior distributions. In other words, we seek to determine whether it is plausible that the available data correspond to an expected outcome generated by our forward simulation framework. To this end, we draw 1000 parameter sets from the posterior distributions and generate mock data vectors using our simulation pipeline in a fiducial area of 1000 deg$^2$. 

Since we do not have access to an analytical likelihood, a traditional goodness-of-fit test cannot be performed to evaluate the efficiency of our trained model. However, to retain this evaluation metric, we implement a modified version of the $\chi^2$ goodness-of-fit test following the approach presented in \cite{vonWietersheim-Kramsta2024}. Namely, we introduce a test statistic $C$ as the Poissonian log-likelihood of the observed data points with respect to the mean of the 1000 simulated samples,

\begin{equation} \label{Eq: T}
C = -2\sum_{\text{bins}} \log \Lambda(\mu,k)
\end{equation}

where $\Lambda(\mu,k)$ represents the Poisson probability distribution in each bin, with $\mu$ being the mean of all realizations and $k$ the value of a specific realization. In the Gaussian case, $C$ naturally translates into the standard $\chi^2$ formula, making it a direct generalization of the traditional $\chi^2$ test to the case of Poisson-distributed data.

Figure \ref{Fig:Gof} shows the cumulative distribution function (CDF) of $C$ values obtained from the 1000 realizations in blue, along with the $C$ value corresponding to the mock data, shown as a red dashed line. This comparison allows us to assess the coverage of the posterior distributions of parameters with respect to the \textit{true} value. We can see that the test statistic value obtained for the data corresponds to a CDF value of 0.51, such that 49\% of the randomly generated samples exhibit a value of the test statistic that is larger than the value obtained for the mock sample. Therefore, the value of the test statistic obtained for the mock samples lies within the range that is expected in case the model is perfectly able to reproduce the mock data and the only source of uncertainty is statistical noise, which is considered acceptable under the standard criterion that p-values greater than 5\% indicate consistency with the model.

\subsection{Posterior distributions} \label{sec: Posterior distributions}
\begin{figure}
\centering
\includegraphics[width=\columnwidth]{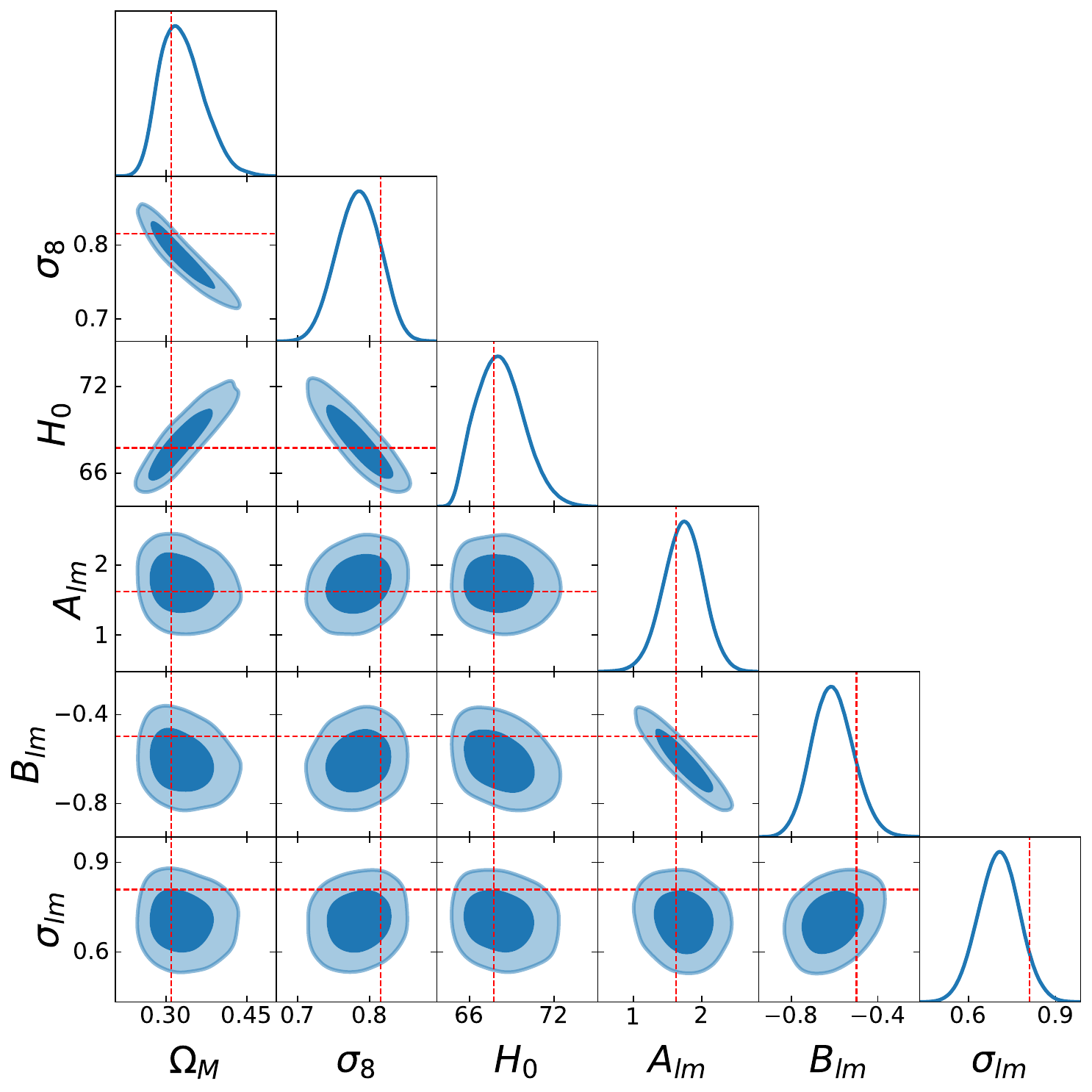}
\caption{Corner plot of the 6 posterior distributions corresponding to each parameter of the pipeline: the 3 cosmological parameters $\Omega_m$, $\sigma_8$, and $H_0$, and the 3 parameters from the M-L scaling relation $A_{lm}$, $B_{lm}$, and $\sigma_{lm}$ corresponding to the slope, the normalization, and the scatter, respectively. The results correspond to a run with flat priors on each parameter, see Table \ref{tab:PriorsSBI}, and a uniform proposal for the training of the model, on a fiducial area of 1000 deg$^2$. The contours represent the 1$\sigma$ and 2$\sigma$ confidence levels and the red dashed lines are the UNIT1i values.}
\label{Fig:corner_flatprop_1000}
\end{figure}

The corner plot shown in Fig. \ref{Fig:corner_flatprop_1000} displays the posterior distributions of the parameters within our pipeline. It includes three cosmological parameters ($\Omega_m$, $\sigma_8$ and $H_0$) and three parameters related to the M-L scaling relation (slope, normalization, and log-normal intrinsic scatter). 

\begin{figure*}
\centering
\centerline{\resizebox{\hsize}{!}
{\includegraphics{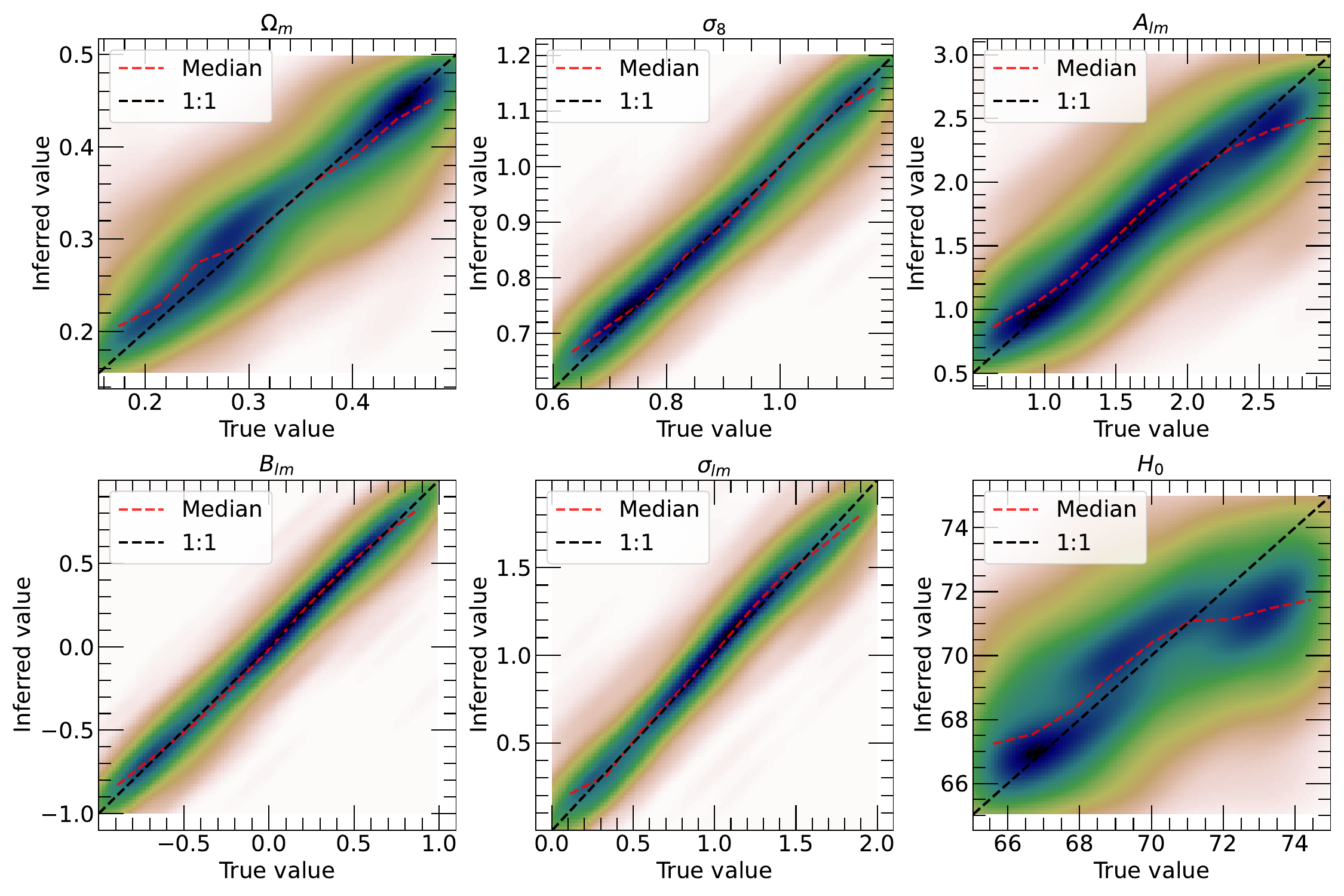}}}
\caption{Results of the coverage test described in Sect. \ref{sec:Coverage test}, comparing the input values (\textit{true value}) to the inferred ones (\textit{inferred value}) for each parameter by taking random points in the posterior distribution chains. The point density is smoothed using a Gaussian kernel density estimator and indicated by the color code, with the median of the points indicated by a red dashed line and the 1:1 line in black.}
\label{Fig:coverage}
\end{figure*}

In Table \ref{tab:parameter results}, we present the best-fitting values obtained with our pipeline, and compare them to the input cosmological parameters used to generate the UNIT1i light cone, as well as to the M–L relation parameters that were fitted directly from the mock catalog. While the cosmological parameters were explicitly set in the simulation, the M–L parameters were not defined a priori, since the mock does not rely on scaling relations to assign luminosities. Instead, we derived them by fitting the mock cluster masses and luminosities. We can see that the retrieved parameters are fully consistent with the input values, which confirms the effectiveness of our approach. The corner plot highlights the known degeneracy between $\Omega_m$ and $\sigma_8$. We also note that $\Omega_m$ and $\sigma_8$ appear to be correlated with $H_0$. The observed degeneracy between $H_0$ and $\Omega_m$ arises from their combined influence on the predicted number of galaxy clusters. While $H_0$ significantly impacts the comoving volume (see Eq. \ref{Eq:Volume}), $\Omega_m$  governs the shape and normalization of the halo mass function, which dictates the number density of massive objects. As a result, a higher $H_0$, which reduces the survey volume, can be compensated by a higher $\Omega_m$ to maintain the same number of detected clusters, and vice versa. This volume–mass function trade-off generates the observed degeneracy. Furthermore, the correlation between $H_0$ and $\sigma_8$ is indirectly driven by their respective degeneracies with $\Omega_m$. Appendix \ref{sec:H0 constraint for different sample size} shows that this effect is not visible in smaller survey areas (e.g., 50 deg$^2$), but becomes significant for larger samples such as the 1000 deg$^2$ case studied here.

A clear correlation is observed between the slope and normalization of the luminosity–mass (M–L) scaling relation, $A_{lm}$ and $B_{lm}$, respectively. We note that the degeneracy between the slope and the normalization of the luminosity-mass relation arises from our choice of pivot point ($10^{14}M_\odot$, see Eq. \ref{Eq:ML}), which is located below the limiting mass considered in the mock analysis. The degeneracy would disappear if the pivot point was set at a slightly higher mass or if the sample extended to lower masses, as is the case, for instance, of the XMM-XXL \citep{Pierre2016} and eROSITA/eRASS1 \citep{Ghirardini2024} samples.

\begin{table*}[t]
    \centering
    \caption{Definition of the 6 pipeline parameters, their corresponding mock UNIT1i values (\textit{true values}), and their inferred values obtained through SBI along with the associated uncertainties. The inferred values were obtained from a run using flat priors for each parameter and a uniform proposal distribution for training the model, performed on a fiducial area of 1000 deg$^2$. The UNIT1i values for $A_{lm}$, $B_{lm}$ and $\sigma_{lm}$ where fitted from the mock.}
    \begin{tabular}{c c c c}
        \hline\hline
        Parameters & Definitions & UNIT1i value & fitted value \\
        \hline
        $\Omega_m$  & Matter density as a fraction of critical density & $0.309$ & $0.330${\raisebox{0.3ex}{\tiny$^{+0.031}_{-0.048}$}}\\
        $\sigma_8$ & Amplitude of density fluctuations on a scale of 8 Mpc/h & $0.815$ & $0.784${\raisebox{0.3ex}{\tiny$^{+0.030}_{-0.030}$}}\\
        $H_0$       & Hubble constant & $67.74$ & $68.2${\raisebox{0.3ex}{\tiny$^{+1.4}_{-1.9}$}}\\
        $A_{lm}$       & Mass-luminosity slope & $1.620$ & $1.730${\raisebox{0.3ex}{\tiny$^{+0.29}_{-0.29}$}}\\
        $B_{lm}$       & Mass-luminosity normalization & $-0.499$ & $-0.608${\raisebox{0.3ex}{\tiny$^{+0.089}_{-0.1}$}}\\
        $\sigma_{lm}$   & Mass-luminosity scatter & $0.809$ & $0.705${\raisebox{0.3ex}{\tiny$^{+0.071}_{-0.071}$}}\\
        \hline
    \end{tabular}
    \vspace{0.5em}
    \label{tab:parameter results}
\end{table*}

\subsection{Coverage test} \label{sec:Coverage test}

To assess the coverage properties of our trained model, we perform a dedicated test aimed at verifying whether the inferred posteriors are statistically consistent with the true parameters. This test is computationally intensive, so we carry it out on the 50 deg$^2$ configuration. We begin by generating catalogs of clusters in a 50 deg$^2$ area, from 10,000 random points in the prior parameter space. For each generated catalog, we apply the model trained on the same survey area and draw 10,000 samples using the \texttt{sbi} framework. From the resulting posterior chains, we then randomly select one point per parameter.

Figure \ref{Fig:coverage} compares the input values (\textit{true value}) to the inferred ones (\textit{inferred value}) for the six parameters of the simulation pipeline. We visualize the results using a Gaussian kernel density estimator and indicate the median of the points with a red dashed line. To minimize the impact of prior boundaries on the inferred parameters, which could bias the results if we adopt the median of the posterior as our point estimate, we instead select a random point from the posterior chain. Nevertheless, we still note a slight overestimation and underestimation of the recovered points close to the lower and upper boundaries of the prior, respectively. This behavior is expected, as the shape of the prior in these regions is highly asymmetric.

We first note that our model effectively constrains the $\sigma_8$ parameter, as evidenced by the high density of points around the 1:1 line. As for $\Omega_m$, the method recovers it in an unbiased manner, with a clear concentration of points around the 1:1 line. However, while the recovery of $\Omega_m$ is accurate, the precision is lower compared to $\sigma_8$, as shown by the larger scatter around the 1:1 line.

Regarding the scaling relation parameters, they are all well constrained, suggesting that our model can constrain simultaneously the cosmological parameters and the M-L scaling relation. While the model precisely constrains the normalization of the M-L relation, the slope is not as well constrained, with the scatter of the points around the 1:1 line showing a larger uncertainty around the true value. This is due to the mass cut imposed by the selection function, which reduces sensitivity to the slope, given the absence of low-mass clusters in the selected sample.

For a 50 deg$^2$ survey $H_0$ is poorly constrained, as the point density is not concentrated around the 1:1 line. Since the cluster count method is not highly sensitive to $H_0$, this outcome is expected. By treating $H_0$ as a free parameter in our simulation pipeline, we marginalize over its uncertainty when constraining the other parameters. Nonetheless, some sensitivity to $H_0$ remains, as the posterior distribution is not completely flat for a 50 deg$^2$ survey. Larger surveys are more sensitive to $H_0$, as demonstrated in the 1000 deg$^2$ case shown in Fig. \ref{Fig:corner_flatprop_1000} and Appendix \ref{sec:H0 constraint for different sample size}.

\section{Discussion} \label{sec:Discussion}

\subsection{Comparison with traditional methods}
\label{sec:Comparison with traditional methods}

Traditional analyses of galaxy cluster counts typically rely on analytic likelihoods and simplified assumptions about the survey selection function and scaling relations. However, such approaches are often unable to marginalize over important systematics, including uncertainties in the observable–mass relation, approximations in the mass function, and incomplete modeling of the selection effects. In contrast, the SBI method used in this work provides a robust alternative by forward-modeling the entire observed cluster population. It fits directly to the observables, without requiring an explicit likelihood. The simulation process naturally accounts for all cosmological dependencies. Crucially, it enables a more accurate treatment of complex effects such as the survey selection function, the intrinsic scatter and covariance in scaling relations, and selection biases linked to profile shapes or morphological properties. It also allows us to marginalize over modeling uncertainties by directly sampling uncertain physical parameters over their allowed range. As a result, it offers a more flexible and potentially more accurate framework for extracting cosmological information from cluster surveys.

Another key advantage of the SBI approach is its ability to integrate multiple datasets into a single pipeline, enabling joint analyses across observables such as clustering, cluster counts, and baryon acoustic oscillations (BAO). While traditional methods can also combine such datasets, SBI naturally accounts for the covariance between observables through the forward modeling process. By predicting multiple observables from the same cosmological model, this approach allows us to enrich the model by adding more data to the observable vector, making the analysis more powerful and flexible. With our framework, it is relatively straightforward to include other cosmological probes to improve the constraints and provide a self-consistent inference across all probes. However, compatibility between probes is not guaranteed, as each may have its own systematic uncertainties. A prudent strategy is to test each probe independently and compare the results. This can reveal potential tensions, opening the door to further investigation. Ultimately, combining probes coherently can yield more robust constraints and a deeper understanding of the cosmological model.

\begin{figure}
\centering
\includegraphics[width=\columnwidth]{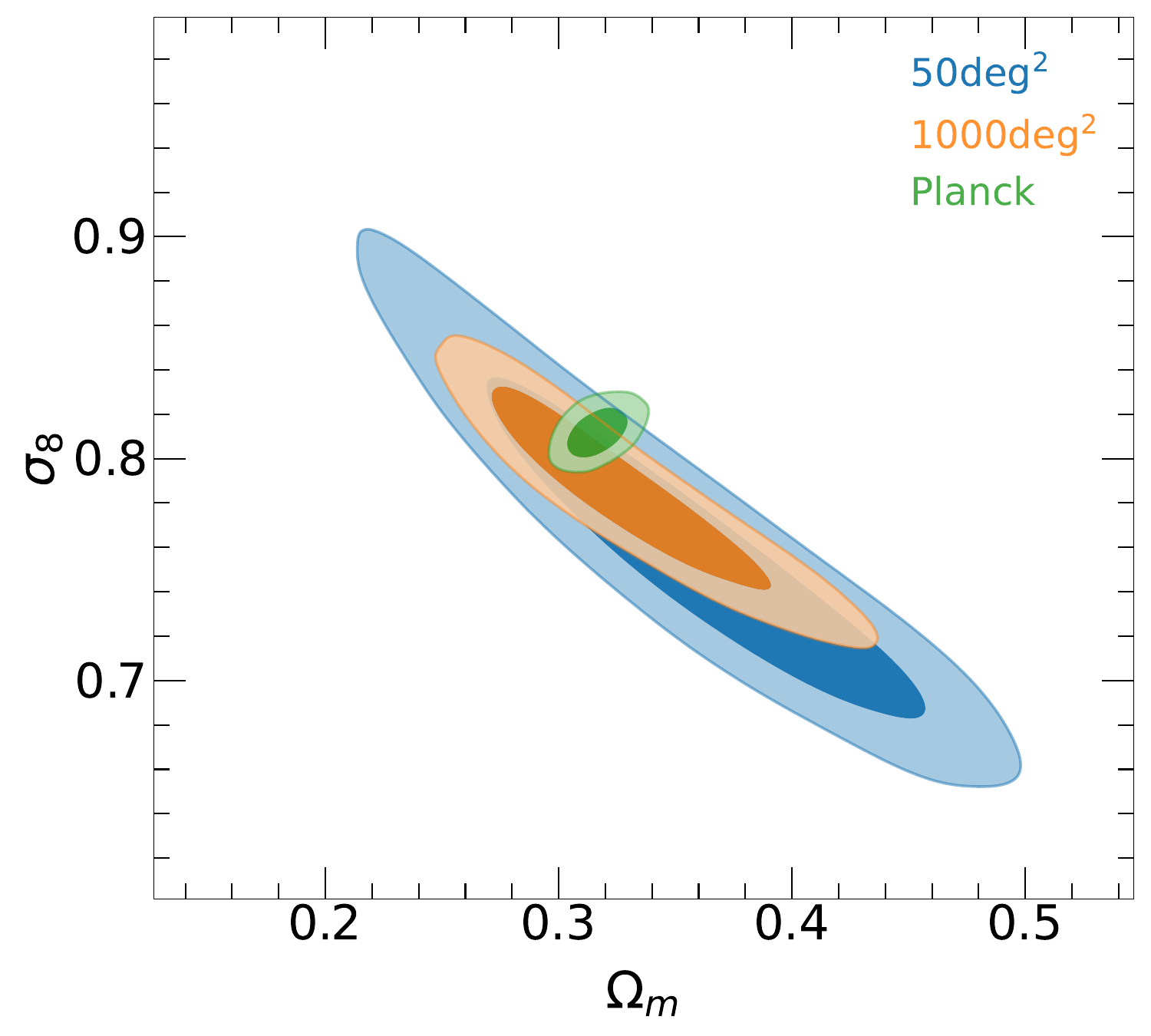}
\caption{Posterior distributions in the $\Omega_m-\sigma_8$ plane for mock survey areas of 50 deg$^2$ in \textit{blue} and 1000 deg$^2$ in \textit{orange}, compared to the Planck values used in UNIT1i in \textit{green}. The contours represent the 1$\sigma$ and 2$\sigma$ confidence levels.}
\label{Fig:50deg_1000deg}
\end{figure}

\subsection{Survey size} \label{sec:Survey size}

\begin{figure*}
\centering
\centerline{\resizebox{\hsize}{!}
{\includegraphics{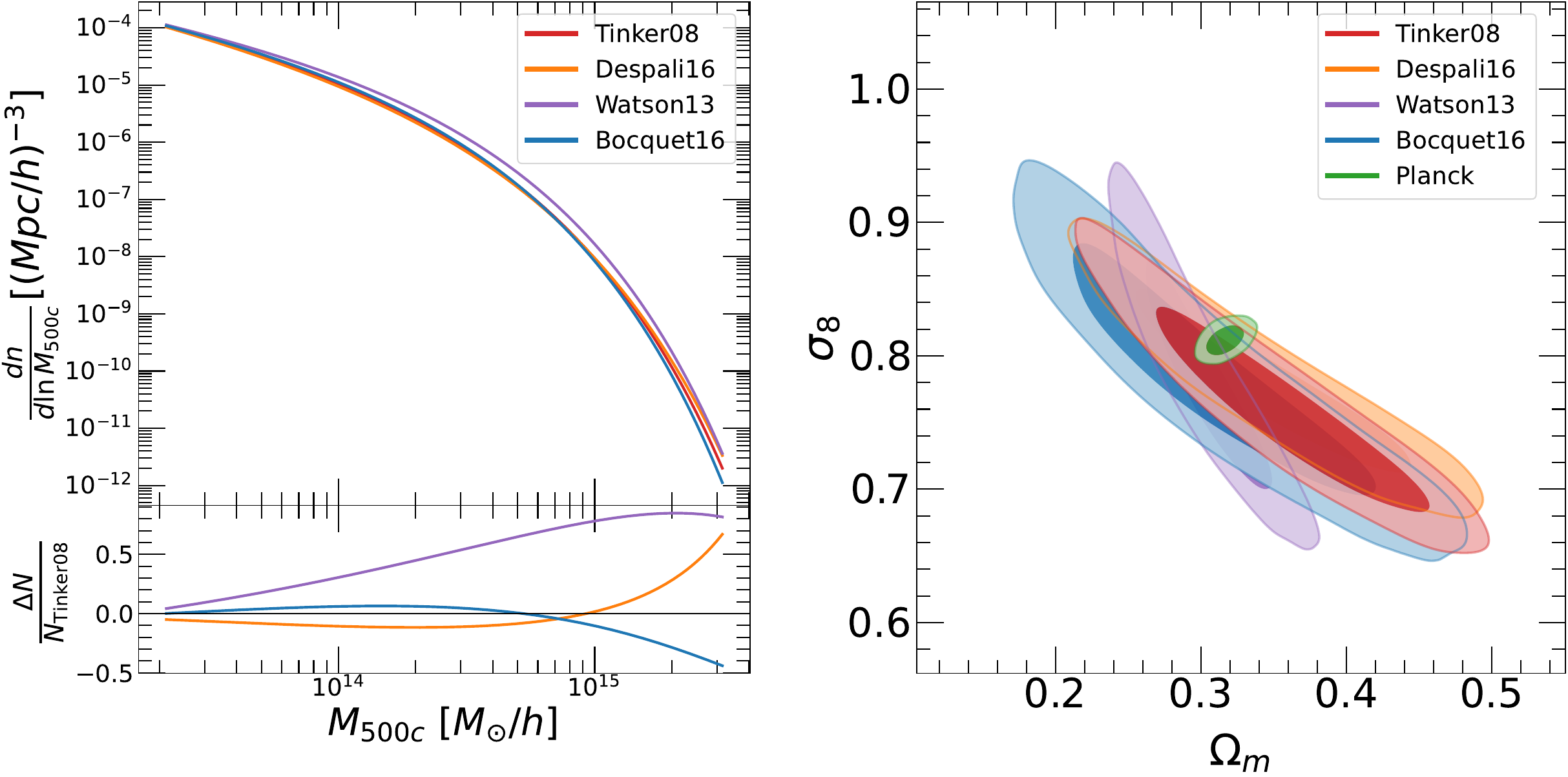}}}
\caption{\textit{left:} Comparison of mass function models available in the \texttt{colossus} Python package relative to the Tinker08 model, in the cosmology of the mock sample, i.e., $\Omega_m = 0.3089, \sigma_8 = 0.8159, H_0 = 67.74$ and for z=0.426, the median redshift of the mock sample. The bottom plot shows the residuals. \textit{right:} Crossed posterior distributions of $\Omega_m$ and $\sigma_8$ for a mock survey covering 50 deg$^2$ with a non-uniform proposal. Different colors represent different runs, each using a distinct mass function model available in the \texttt{colossus} Python package. The Planck value of $\Omega_m$ and $\sigma_8$, which corresponds to the true values implemented in the mock, is also shown for reference.  The contours represent the 1$\sigma$ and 2$\sigma$ confidence levels.}
\label{Fig:Mass_fct_comparison}
\end{figure*}

Here, we investigate the impact of the survey size on the uncertainty of the inferred parameters. To do so, we compare the results obtained from a mock survey covering 50 deg$^2$ with those from a mock survey covering 1000 deg$^2$, focusing on the joint posterior distributions of $\Omega_m$ and $\sigma_8$, see Fig. \ref{Fig:50deg_1000deg}.

In both cases, the parameter estimates converge to values consistent with the Planck results \citep{PlanckCollaboration2016} within 1$\sigma$, confirming that the model reliably recovers cosmological parameters. As expected, increasing the survey size significantly reduces the uncertainties on each parameter, highlighting the advantage of larger datasets in improving parameter constraints. As a step in this direction, we conducted a mock analysis with the same number of clusters as eRASS1 to explore its potential application, see Appendix \ref{sec:eRASS1}.

\subsection{Systematic uncertainties} \label{sec:systematic error}

Here, we analyze the impact of systematic errors in our method and how they affect the results. Specifically, we consider three key sources of systematics: the choice of the mass function, the normalization of the M-T relation, and sample variance. Variations in the mass function can lead to differences in the inferred cosmological parameters, while uncertainties in the M-T normalization directly influence mass estimates and, consequently, constraints on $\sigma_8$ and $\Omega_m$. Finally, sample variance, inherent to the limited survey area, introduces additional fluctuations that must be taken into account in our analysis.

\subsubsection{Mass function model comparison} \label{sec:Mass function model comparison}

For the galaxy cluster count approach, the choice of the mass function introduces a systematic uncertainty. To investigate the impact of the choice of the mass function on the recovered cosmological parameters, we compared four different mass function models implemented in the \texttt{colossus} Python package: Watson13 \citep{Watson2013}, Despali16 \citep{Despali2016}, Tinker08 \citep{Tinker2008}, and Bocquet16 \citep{Bocquet2016}. The first three are calibrated using dark matter only N-body simulations, whereas Bocquet16 is based on hydrodynamical simulations that incorporate baryonic physics. 

In the left-hand panel of Fig. \ref{Fig:Mass_fct_comparison} we plot the four mass function models evaluated at the median redshift of the mock sample ($z=0.426$) and display the ratio of the various models to our default model (Tinker08). In the mass range covered by our data, the Tinker08, Despali16 and Bocquet16 models are consistent with one another at the $\sim20\%$ level, whereas the Watson13 model deviates by more than 50\%. Large differences between the models can be observed at very high masses ($>10^{15}M_\odot$). However, such very massive clusters are very rare and mostly absent within an area of 50 deg$^2$. In the right-hand panel of Fig. \ref{Fig:Mass_fct_comparison} we show the values of $\Omega_m$ and $\sigma_8$ recovered with our pipeline over a 50 deg$^2$ area for the four different mass functions. The systematic uncertainty induced by the choice of the mass function is not particularly problematic over this survey size, similar to the XMM-XXL field. We adopt the Tinker08 model by default as it was found to provide the best match to the intrinsic UNIT1i mass function \citep{Seppi2022}. The largest deviations compared to the default Tinker08 model reach up to $\sim16\%$ for $\Omega_m$ and $\sim4\%$ for $\sigma_8$, with the Watson13 model showing both the largest offset and a noticeably different degeneracy direction between $\Omega_m$ and $\sigma_8$, which is due to the different slope of the HMF in the $10^{14}-10^{15}M_\odot$ range compared to the other models.

This effect is expected to be more relevant for larger fields such as eROSITA. For a 50 deg$^2$ area, the statistical uncertainties dominate and mask the systematic differences due to the mass function choice. While the systematic uncertainty from the mass function will eventually become dominant as survey size increases, it remains subdominant compared to statistical errors at the scale of our current analysis. In other words, as survey areas grow and statistical uncertainties shrink, the choice of mass function will play an increasingly important role in the error budget.

\subsubsection{Sample variance} \label{sec:sample variance quantification R}
\begin{figure*}
\centering
\centerline{\resizebox{\hsize}{!}{\includegraphics{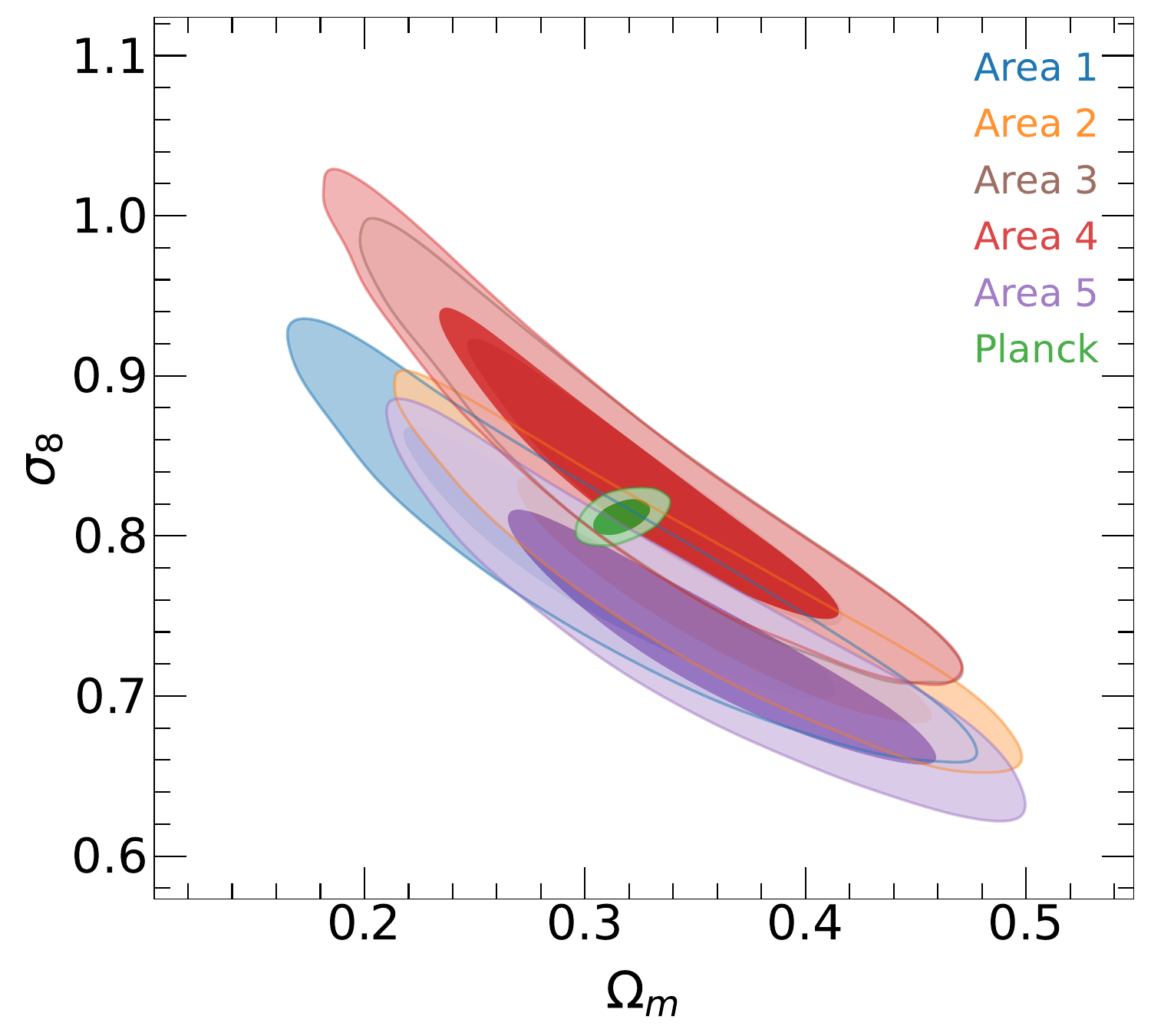}\includegraphics{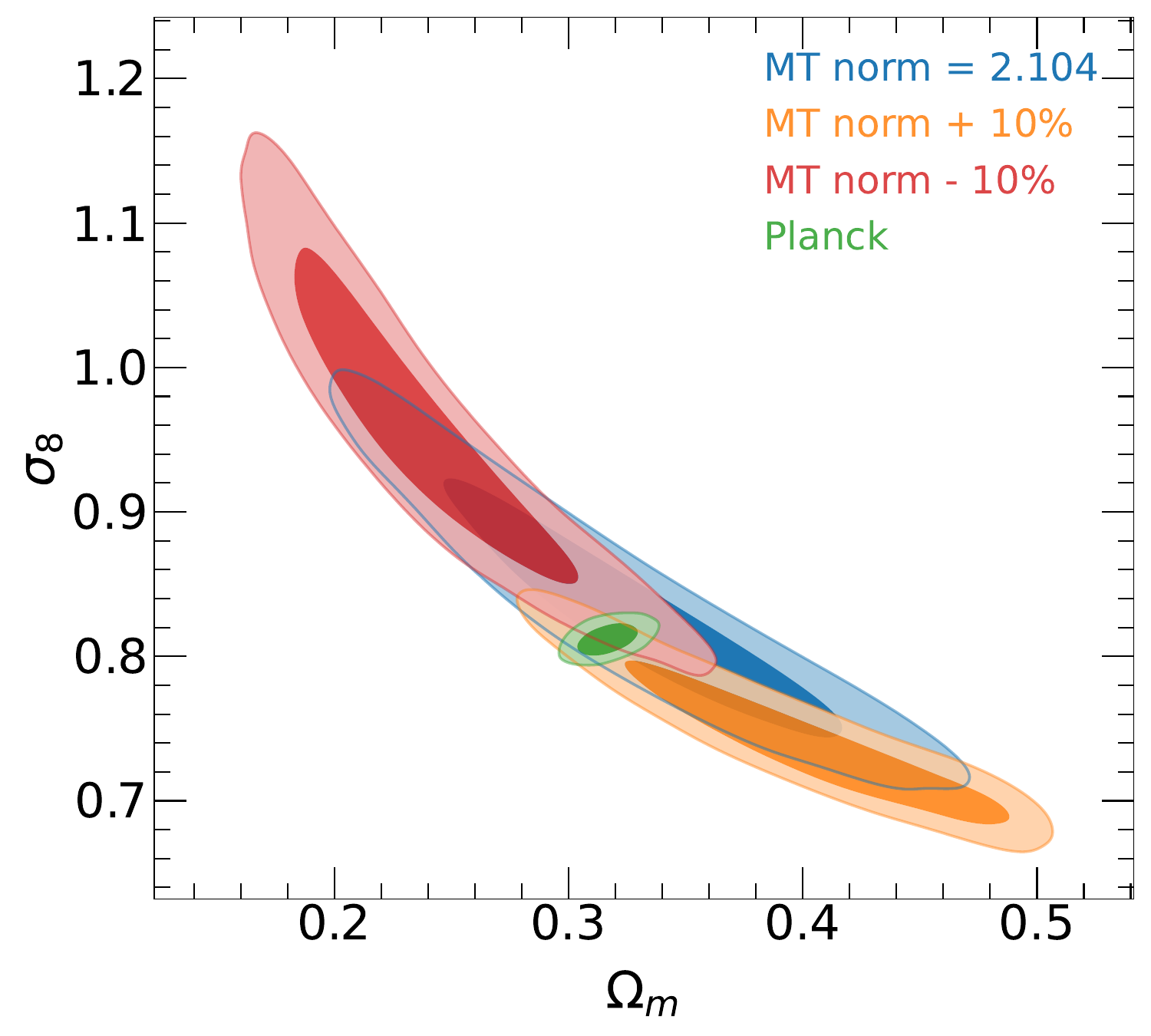}}}

\caption{\textit{Left:} Sample variance test in the form of 2D posterior distributions of $\Omega_m$ and $\sigma_8$ for different 50 deg$^2$ regions of the mock sample with a uniform proposal. The contours represent the 1$\sigma$ and 2$\sigma$ confidence levels. \textit{Right:} Crossed posterior distributions of $\Omega_m$ and $\sigma_8$ for a mock survey covering an area of 50 deg$^2$ with a uniform proposal. The distribution shown in \textit{blue} assumes a M-T normalization parameter of $2.104$ (see \ref{sec:Scaling relation}), while the distributions in \textit{orange} and \textit{red} correspond to variations of plus and minus 10\% around this value, respectively. The results are compared with the Planck values used in UNIT1i, represented in \textit{green}. The contours indicate the 1$\sigma$ and 2$\sigma$ confidence intervals.}
\label{Fig:sample_variance_MT}
\end{figure*}

We investigated the impact of sample variance by retrieving the cosmological parameters from five different areas extracted from the UNIT1i light cone, each covering 50 deg$^2$, see the left-hand panel of Fig.\ref{Fig:sample_variance_MT}. This sample variance, which arises from large-scale structure and represents a source of error not accounted for in our simulation pipeline, could potentially affect our results. However, the differences observed between the various tested regions are smaller than the associated error bars. Therefore, we can reasonably conclude that the sample variance does not significantly impact the results for the 50 deg$^2$ areas and can be considered negligible in this context. Sample variance decreases with survey size and is thus not expected to play an important role for wider surveys.

\subsubsection{Mass-Temperature normalization} \label{sec:Mass-Temperature normalization}

The information on the mass scale in our simulation pipeline is encoded within the normalization of the M-T relation, which we set using external constraints. The M-T normalization parameter is fixed at 2.104 keV at the pivot mass of $10^{14}M_\odot$ (see Section \ref{sec:Scaling relation}), a value calibrated from WL measurements that carries an estimated uncertainty of about 10\%. This external uncertainty could potentially introduce a systematic bias on the resulting cosmological parameters. To assess the impact of the mass calibration, we test how varying this parameter by $\pm10\%$ affects our results. The right-hand panel of Fig. \ref{Fig:sample_variance_MT} shows the posterior distributions in the $\Omega_m-\sigma_8$ plane when changing the normalization of the M-T relation by $\pm10\%$, over an area of 50 deg$^2$ area. We repeated the analysis over the same five regions as for the sample variance (see Section \ref{sec:sample variance quantification R}) and display here a single one as a representative example. Similar trends can be seen in all five mocks areas we tested. We can see that the contours shift as a function of the M-T normalization. When the M-T normalization increases by $10\%$, $\Omega_m$ goes up by $17\%$ with a median value of $0.384$ while $\sigma_8$ drops by $5\%$ with a median value of $0.736$. On the other hand, decreasing the M-T normalization by $10\%$ leads to a $28\%$ drop in $\Omega_m$ with a median value of $0.238$ and a $11\%$ increase in $\sigma_8$ with a median value of $0.861$. Interestingly, we observe that while both $\Omega_m$ and $\sigma_8$ change significantly, the quantity $S_8 = \sigma_8 (\Omega_m / 0.3)^{0.3}$ remains almost unchanged, varying by only $4\%$. Indeed, the contours move along the well-known degeneracy between these parameters, such that the value of $S_8$ remains almost unchanged. Therefore, our analysis shows that the value of $S_8$ inferred with our method is robust against uncertainties in the absolute mass calibration. The retrieved values of $\Omega_m$, $\sigma_8$, and $S_8$ over the five separate 50 deg$^2$ are provided in Appendix \ref{sec:MT normalization effect}.

\section{Conclusion} \label{sec:Conclusion}

In this work, we have demonstrated the effectiveness of SBI as a powerful method for cosmological analysis. This approach enables us to forward-fit observables directly, incorporating all cosmological dependencies naturally into the simulation pipeline. Generating observables directly allow for the precise application of the selection function within the parameter space where it is defined, ensuring robust and accurate results. Additionally, the implementation of a covariance matrix as a proposal during the training phase of the model significantly accelerates convergence, making the method more efficient and well-suited to this type of problem.

To validate our method, we used mock cluster samples extracted from the UNIT1i dark matter halos simulation. The N-body halos were assigned X-ray properties (temperature and luminosity) using a semi-analytic approach \citep{Comparat2020}, and mock catalogs were generated from the resulting light cone over areas of 50 deg$^2$ and 1000 deg$^2$. We first applied our method to the 1000 deg$^2$ mock sample, and confirmed that the method performs as expected when comparing the retrieved cosmological parameters with the values that were used to generate the N-body simulation. Additionally, we verified that increasing the survey size significantly improves the precision of parameter estimates, highlighting the potential of this approach for future large-area surveys.

Furthermore, we assessed the impact of systematic errors related to sample variance and the choice of the mass function model. We found that for an area of 50 deg$^2$, these systematic uncertainties are subdominant, further validating the robustness of the method for this scale. However, these systematic uncertainties can become substantial in cluster count experiments covering wider areas, such as \emph{eROSITA} \citep{Ghirardini2024} or \emph{Euclid} \citep{Bhargava2025}.

One key factor that significantly improves the precision of the $\sigma_8$ parameter is the inclusion of information on the temperature distribution. Since the temperature acts as an effective proxy for the halo mass, it provides tighter constraints on the mass distribution, which, in turn, enhances our ability to accurately infer $\sigma_8$. This demonstrates the importance of low-scatter mass proxies for more precise cosmological parameter estimates.

\begin{acknowledgements}
We thank Benjamin Joachimi, Michael Burgess, Ben Maughan, and Sebastian Grandis for useful discussions that lead to significant improvements in the paper. MR, DE, and RS acknowledge support from the Swiss National Science Foundation (SNSF) through grant agreement \#200021\_212576. K.U. acknowledges support from the National Science and Technology Council, Taiwan (grant NSTC 112-2112-M-001-027-MY3) and the Academia Sinica Investigator award (grant AS-IA-112-M04).
\end{acknowledgements}

\newpage
\bibliographystyle{aa} % style aa.bst
\bibliography{main}

\newpage
\appendix

\section{Mock selection impact} \label{sec:Mock selection impact}

\begin{figure*}
\centering
\includegraphics[width=\columnwidth]{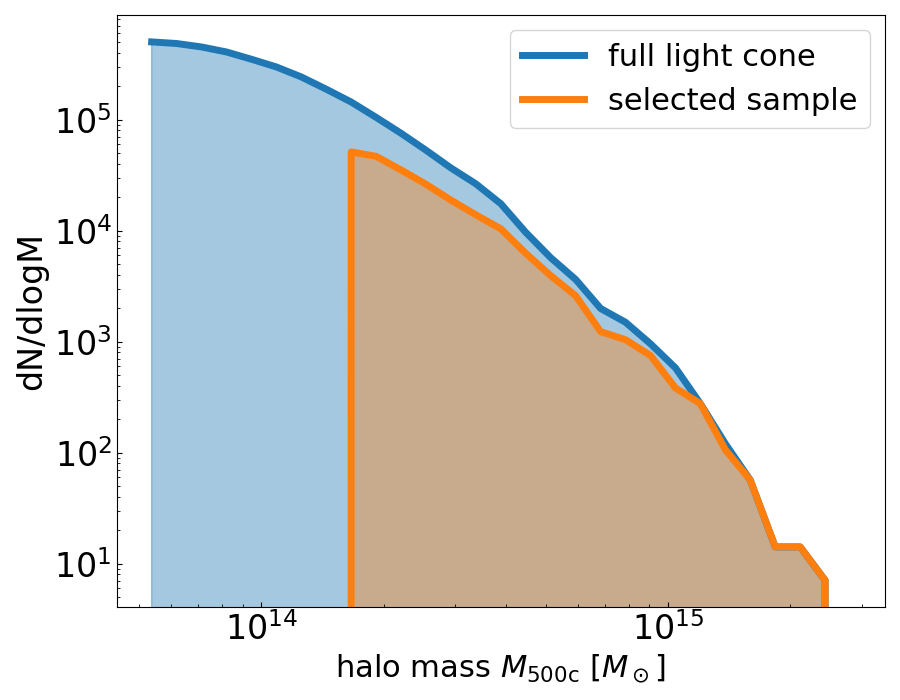}
\includegraphics[width=\columnwidth]{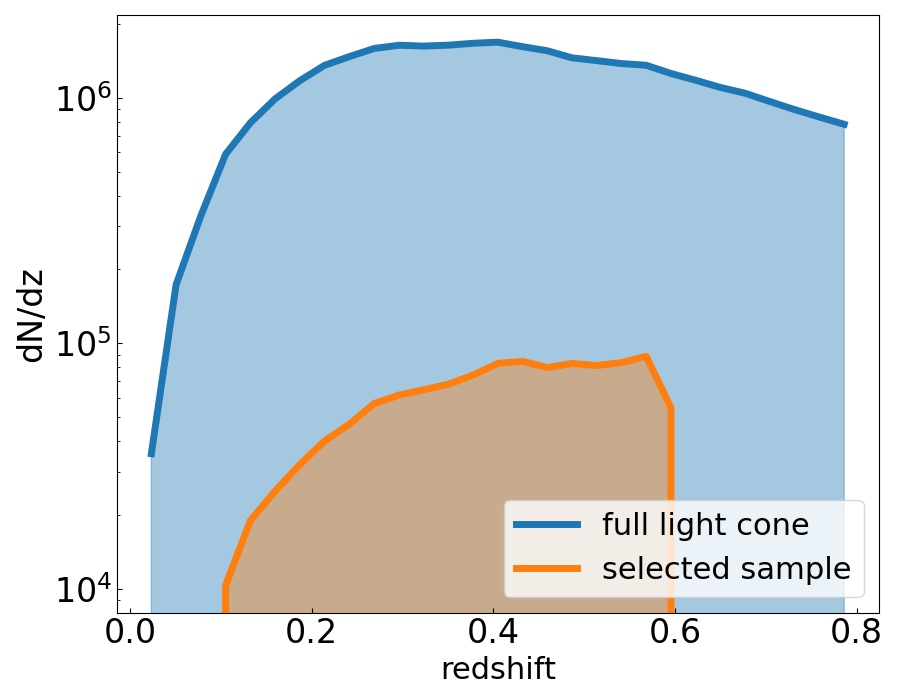}
\caption{Comparison between the complete cluster population in the simulation from \cite{Seppi2022} and the selected one. \textbf{Left-hand panel}: Number of clusters per logarithmic unit of halo mass. \textbf{Right-hand panel}: Number of clusters per unit of redshift. The blue color refers to the full population, the orange to the one after the application of the ideal selection function.}
\label{Fig:Nz_Nm_mock}
\end{figure*}

We show how the selection impacts the overall mass and redshift distribution in Fig. \ref{Fig:Nz_Nm_mock}. The left-hand (right-hand) panel displays the halo population per unit mass (redshift). The blue color identifies the full cluster population, the orange one encodes the ideal selection function. The loss of low mass groups makes our selected sample incomplete as a function of redshift. Excluding the low redshift window below $z<0.1$ causes the loss of some massive clusters. This is not an issue because it is easily accounted by the selection function. 

\section{Core radius distribution} \label{sec:core radius distribution}

\begin{figure}[H]
\centering
\includegraphics[width=\columnwidth]{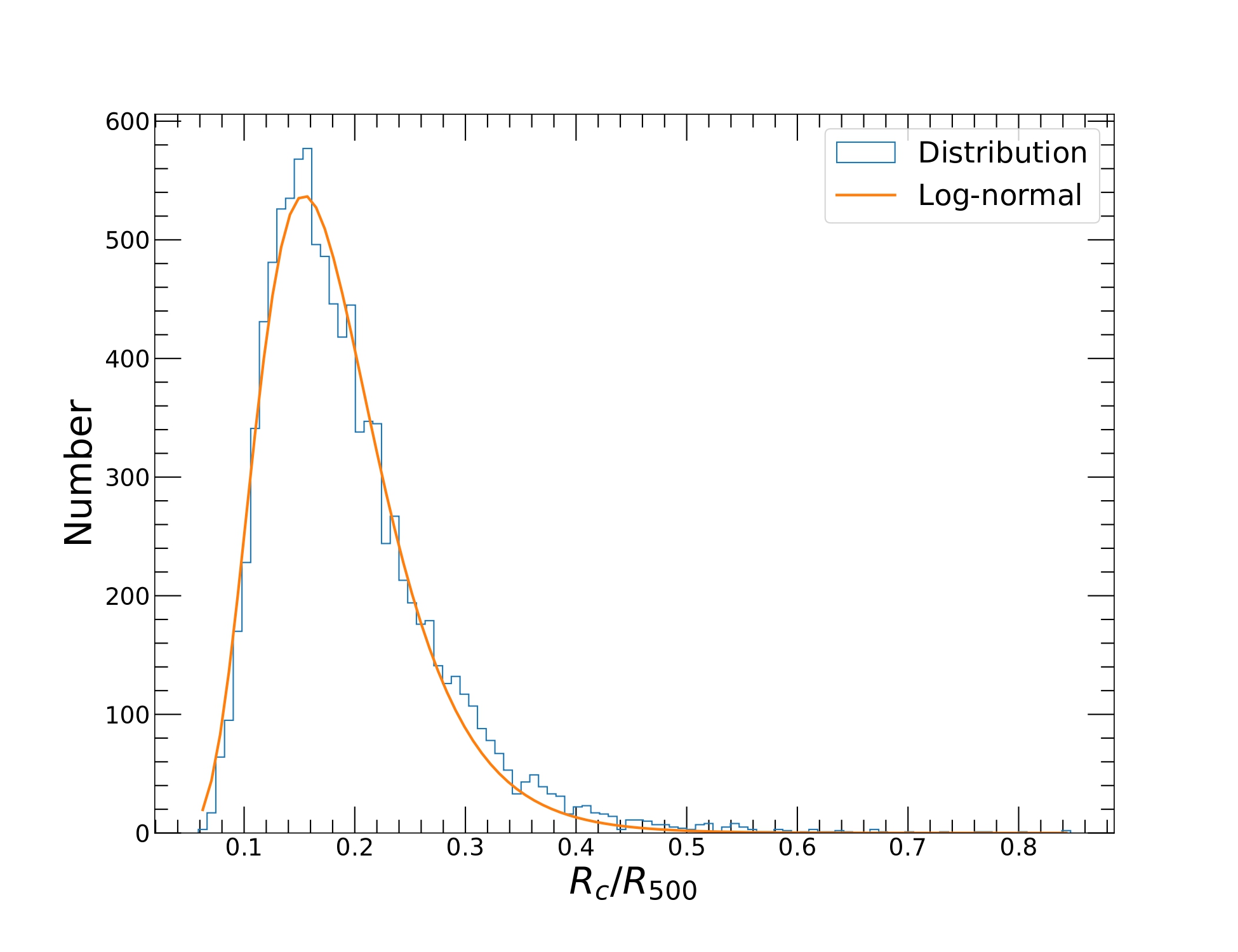}
\caption{Distribution of scaled core radii extracted from 10,000 surface brightness distributions generated with the \citet{Comparat2020} model. The distribution (blue) can be accurately represented with a log-normal distribution with $\mu=0.143$ and $\sigma=0.332$.}
\label{Fig:RcR500}
\end{figure}

The gas distribution model introduced in Sect. 3.6 relies on the prior knowledge of the core radius distribution, since the shape of the generated profiles is parameterized with a single parameter $R_c$ (Eq. 10). To model the core radius distribution, we use the halo gas distribution model of \citet[see Sect. \ref{sec:Mock sample description}]{Comparat2020} to generate a library of emission measure profiles, and we fit the generated profiles with the model given in Eq. 10. We then study the distribution of the scaled core radius, $R_c/R_{500}$. We find that the scaled core radius distribution can be well described by a lognormal distribution with $\mu=0.143$ and $\sigma=0.332$ (see Fig. \ref{Fig:RcR500}). The resulting distribution is used to draw a value of $R_c$ for each simulated system, accounting for the covariance between $R_c$ and $L_x$ \citep{Käfer2019}.

\section{Impact of temperature} \label{sec:Impact of temperature}

\begin{figure}[H]
\centering
\includegraphics[width=\columnwidth]{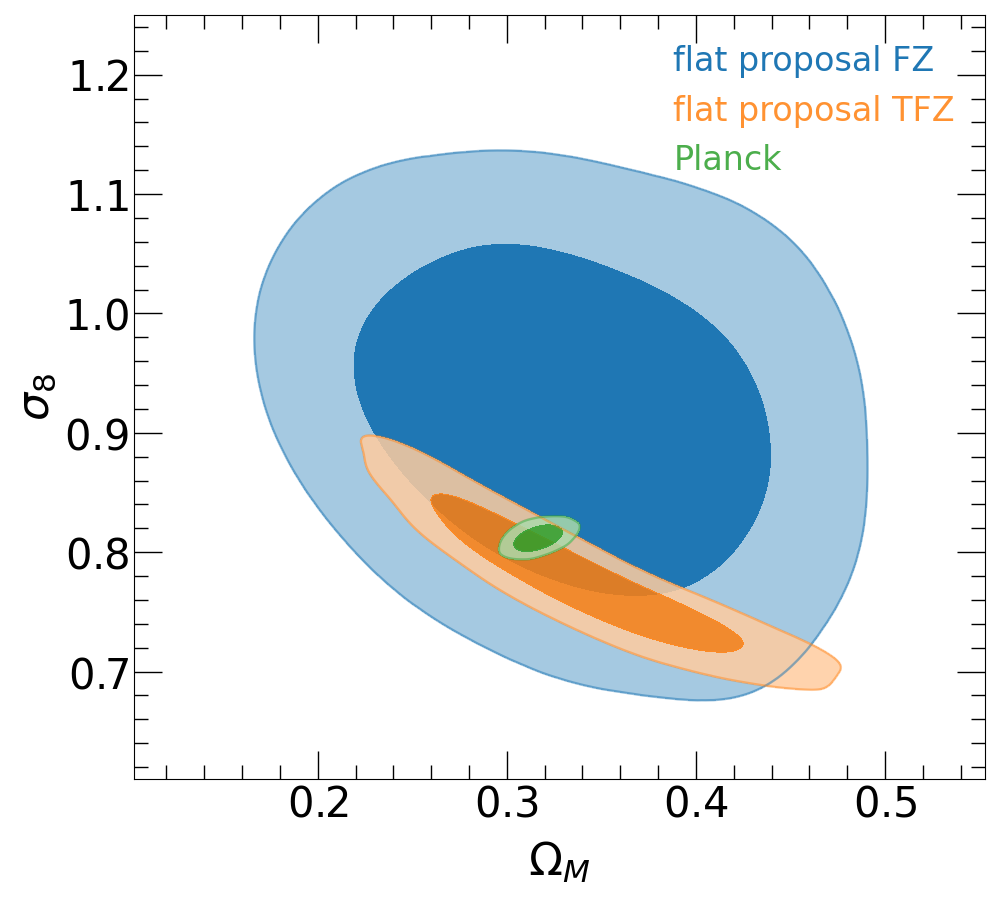}
\caption{Crossed posterior distributions of $\Omega_m$ and $\sigma_8$ for a mock survey area of 50 deg$^2$ with a uniform proposal. The distribution in \textit{blue} considers only the flux and redshift distribution, while the one in \textit{orange} also includes the temperature distribution. The results are compared to the Planck values used in UNIT1i in \textit{green}. The contours represent the 1$\sigma$ and 2$\sigma$ confidence levels.}
\label{Fig:FZ_TFZ}
\end{figure}

Most analyses of cluster populations typically model only the flux and redshift distributions, as the flux is the most direct observable quantity in an X-ray survey \citep{Mantz2015a, Pacaud2018, Ghirardini2024}. However, the temperature is known to be an accurate indicator of the cluster mass \citep{Nagai2007, Pop2022, Braspenning2024}. Relying on flux alone provides no constraint on the mass calibration, which limits the cosmological information that can be extracted. Since the M–T scaling relation is externally calibrated using weak lensing, incorporating the temperature distribution, as done in this work, acts as a mass calibration and leads to significantly improved constraints on cosmological parameters. To evaluate its impact, we applied our method both with and without temperature information and compared the resulting parameter constraints. As shown in Fig. \ref{Fig:FZ_TFZ}, including temperature helps to break the degeneracy between $\Omega_m$ and $\sigma_8$, leading to more precise and distinct parameter estimates. The HMF is highly sensitive to $\sigma_8$, particularly at the high-mass end. Therefore, having a more precise constraint on the mass distribution greatly improves the accuracy of the $\sigma_8$ constraint.

\section{$H_0$ constraint for different sample size} \label{sec:H0 constraint for different sample size}

\begin{figure*}
\centering
\centerline{\resizebox{\hsize}{!}
{\includegraphics{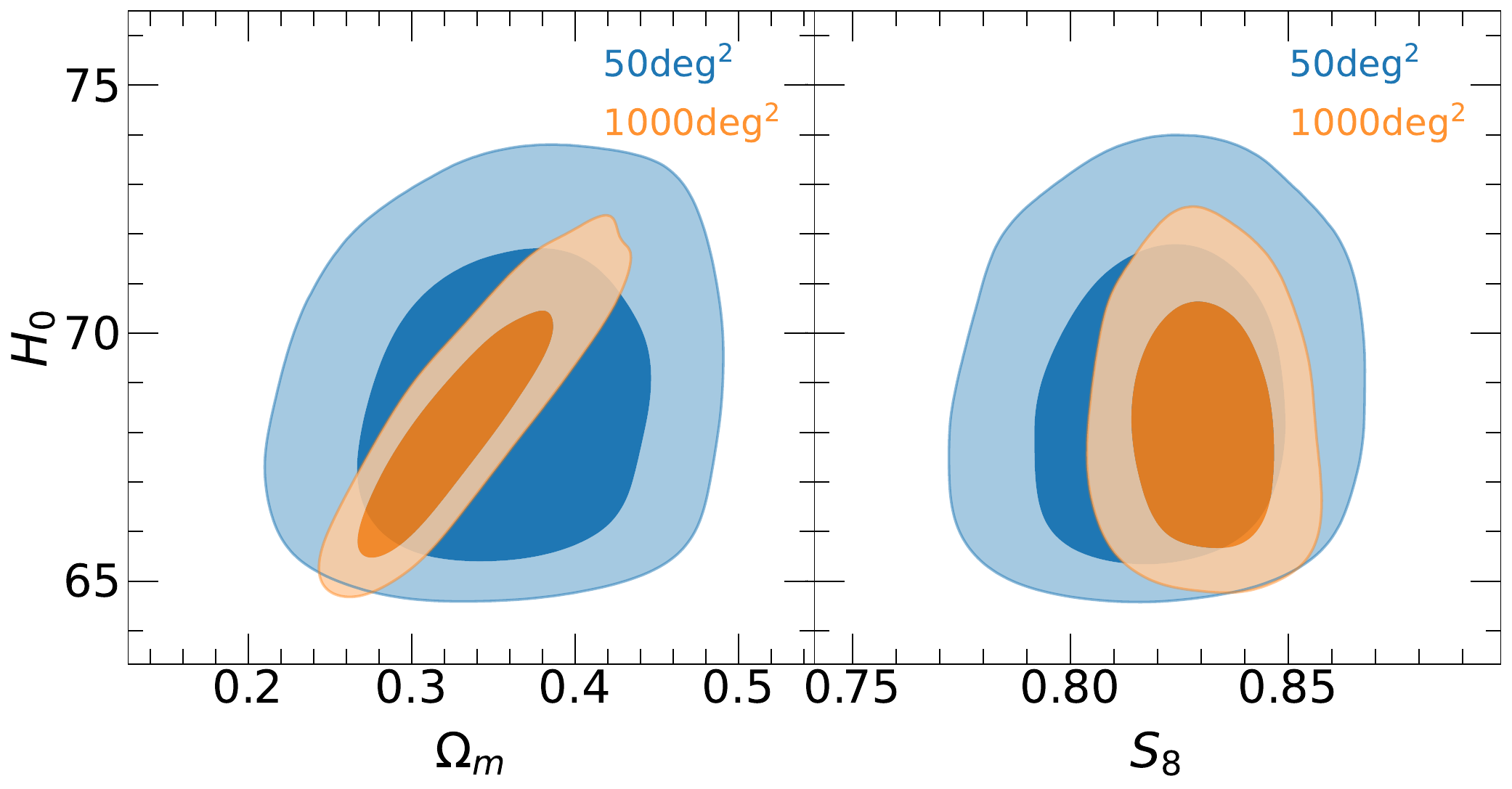}}}
\caption{Crossed posterior distributions of $H_0$ and $\Omega_m$ (on the left) and $H_0$ and $S_8$ (on the right) for mock surveys covering areas of 50 deg$^2$ in \textit{blue} and 1000 deg$^2$ in \textit{orange} , with a uniform proposal. The contours indicate the 1$\sigma$ and 2$\sigma$ confidence intervals.}
\label{Fig:H0_Om0_size}
\end{figure*}

\section{M-T normalization effect} \label{sec:MT normalization effect}

\begin{figure*}
\centering
\centerline{\resizebox{\hsize}{!}
{\includegraphics{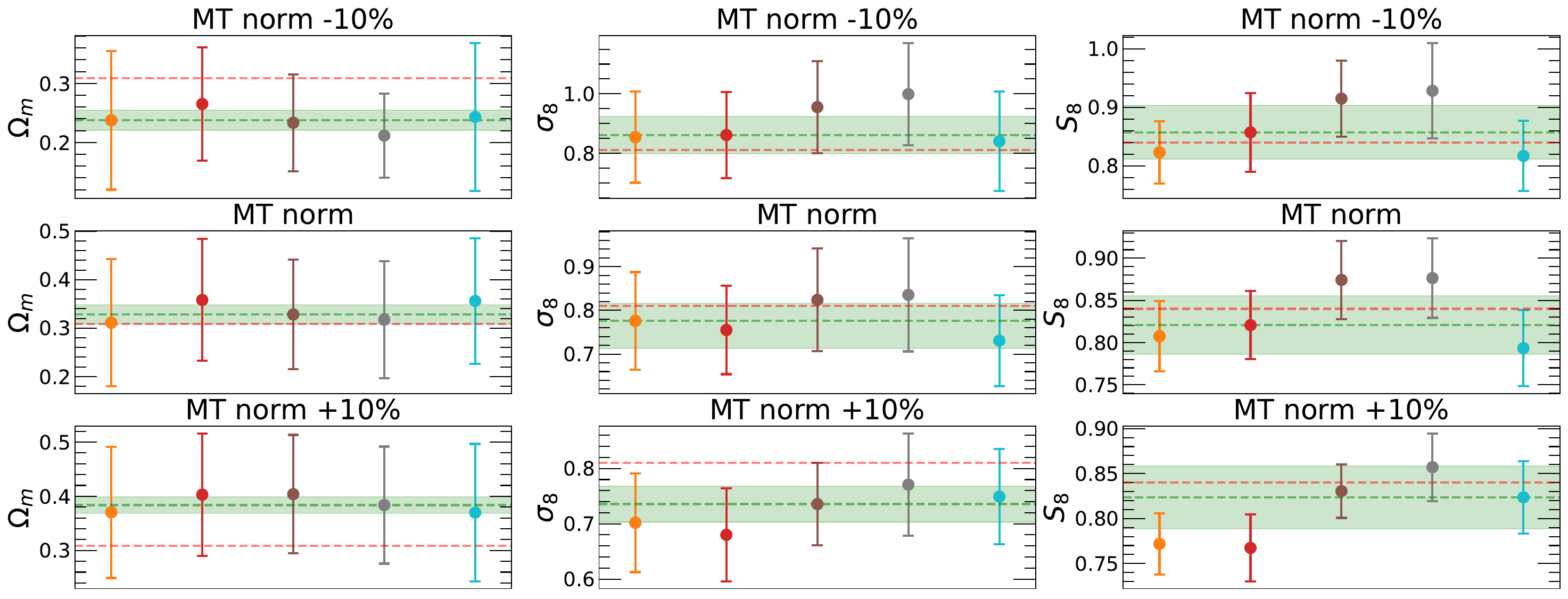}}}
\caption{Impact of the M-T normalization parameter on the inferred cosmological parameters $\Omega_m$, $\sigma_8$, and $S_8$. The second row corresponds to a normalization value of 2.104, while the first and third rows represent a 10\% decrease and increase in normalization, respectively. Each column corresponds to a different cosmological parameter: $\Omega_m$, $\sigma_8$, and $S_8$. Within each plot, the five different 50 deg$^2$ regions of the mock sample are displayed. The red dashed line indicates the true value implemented in the mock, while the green dashed line represents the median of the five inferred values, with the shaded green region denoting the variance.}
\label{Fig:Om0_sig8_s8_5areas_MTnormpm}
\end{figure*}

\section{Fisher proposal values} \label{sec:Fisher proposal}

\begin{figure}[H]
\centering
\includegraphics[width=\columnwidth]{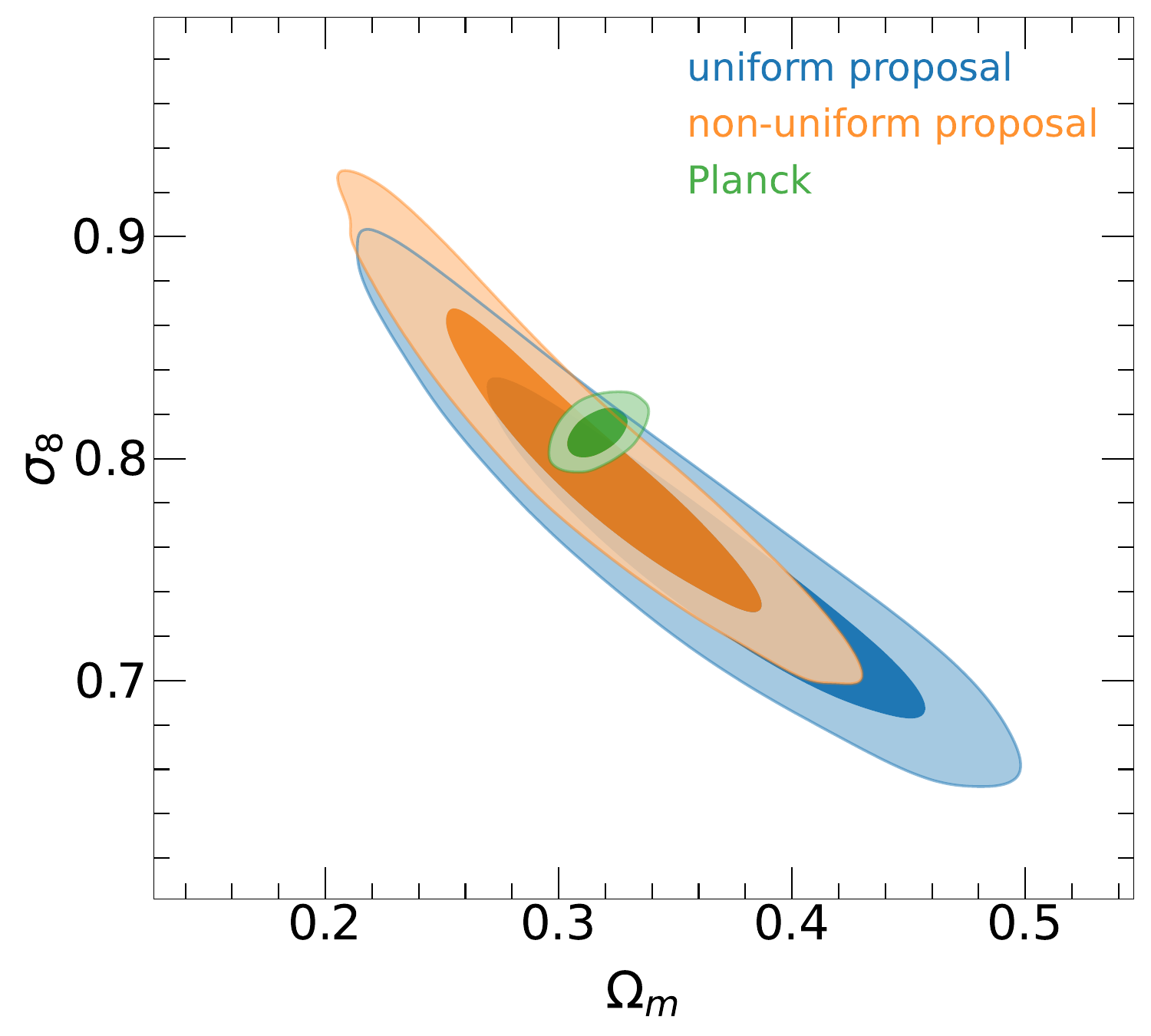}
\caption{Crossed posterior distributions of $\Omega_m$ and $\sigma_8$ in the case of a uniform proposal and a non-uniform proposal informed by a Fisher matrix along with the parameter medians and boundaries for the SBI training, on 50 deg$^2$. The contours represent the 1$\sigma$ and 2$\sigma$ confidence levels.}
\label{Fig:proposal_test_2}
\end{figure}

The standard approach for SBI to train our model involves exploring the parameter space uniformly across its entire range. This parameter space comes from prior distributions that are defined based on our knowledge or assumptions about the possible values for each parameter. To make this process more efficient and avoid loosing time by exploring useless regions of the parameter space, we adjusted the training by instructing SBI to refine the grid of simulations using a non-uniform proposal. This proposal is a multivariate normal distribution informed by a Fisher matrix along with the parameter medians and boundaries. These values are derived from 100,000 posterior distribution points based on the 1000 deg$^2$ run with flat priors and a uniform proposal, see Tables \ref{tab:MinMaxCovSBI} and \ref{tab:fisher_matrix}.
Since statistical error and cosmic variance depend on the survey area, the variance of each parameter must be scaled accordingly when applying this proposal to a different area. For instance, to adapt the 1000 deg$^2$ proposal to a 50 deg$^2$ run, the variance should be adjusted by a factor of 20. Figure \ref{Fig:proposal_test_2} is a comparison of the result obtained using a uniform proposal with those from a non-uniform proposal in terms of the joint posterior distributions of $\Omega_m$ and $\sigma_8$. 

Both configurations converge to parameters consistent with the Planck values used in the mock, confirming that the method reliably recovers cosmological parameters in both cases. However, the non-uniform proposal achieves faster convergence to the true parameter values compared to the flat proposal which is computationally interesting for more parameters. Although this improvement depends on the initial run used to calculate the Fisher matrix. If the model is modified, the Fisher matrix must be recalculated, which implies that the process may not be as quick for subsequent iterations.

\begin{table}[H]
\centering
\caption{Minimum, maximum and median of each parameter derived from 100’000
points of the posterior distribution of 10’000 simulations in a mock area of 1000 deg$^2$. These values are used for the non-uniform proposal distribution for training the model with MCMC. All values are rounded to three decimal places.}
\begin{tabular}{c c c}
\hline\hline
Parameter & (min, median, max)\\
\hline
$\Omega_m$ & (0.155, 0.326, 0.5) \\
$\sigma_8$ & (0.6, 0.785, 1.2)\\
$H_0$ & (65, 68.111, 75)\\
$A_{lm}$ & (0.5, 1.733, 3)\\
$B_{lm}$ & (-1, -0.611, 1)\\
$\sigma_{lm}$ & (0, 0.706, 2)\\
\hline
\end{tabular}
\vspace{0.5em}
\label{tab:MinMaxCovSBI}
\end{table}

\begin{table}[H]
    \centering
    \caption{Fisher matrix with values derived from 100'000 points of the posterior distribution of 10'000 simulations in a mock area of 1000 deg$^2$. As the covariance matrix comes from a run on 1000 deg$^2$, there is here a factor 20 on each value to make it usable on 50 deg$^2$. All values are rounded to three decimal places.}
    \begin{tabular}{ccccccc}
            \hline\hline
            & $\Omega_m$ & $\sigma_8$  & $H_0$ & $A_{lm}$ & $B_{lm}$ & $\sigma_{lm}$\\
            \hline
            $\Omega_m$  &  0.0322  & -0.0221 & 1.1795 & -0.0368 & -0.0149 & -0.0035 \\
            $\sigma_8$  & -0.0221 &  0.0175 & -0.8491&  0.0339 &  0.0092 &  0.0052  \\
            $H_0$       &  1.1795 & -0.8491 & 55.2760 & -0.4603 & -0.9532 & -0.2822  \\
            $A_{lm}$    & -0.0368 &  0.0339 & -0.4603 &  1.7106 & -0.5067 & -0.0552  \\
            $B_{lm}$    & -0.0149 &  0.0092 & -0.9532 & -0.5067 &  0.1831 &  0.0388  \\
            $\sigma_{lm}$ & -0.0035 &  0.0052 & -0.2822 & -0.0552 &  0.0388 & 0.1006  \\
            \hline
    \end{tabular}
    \vspace{0.5em}
    \label{tab:fisher_matrix}
\end{table}

\section{Sbi configurations} \label{sec:Sbi configurations}

\begin{figure}[H]
\centering
\includegraphics[width=\columnwidth]{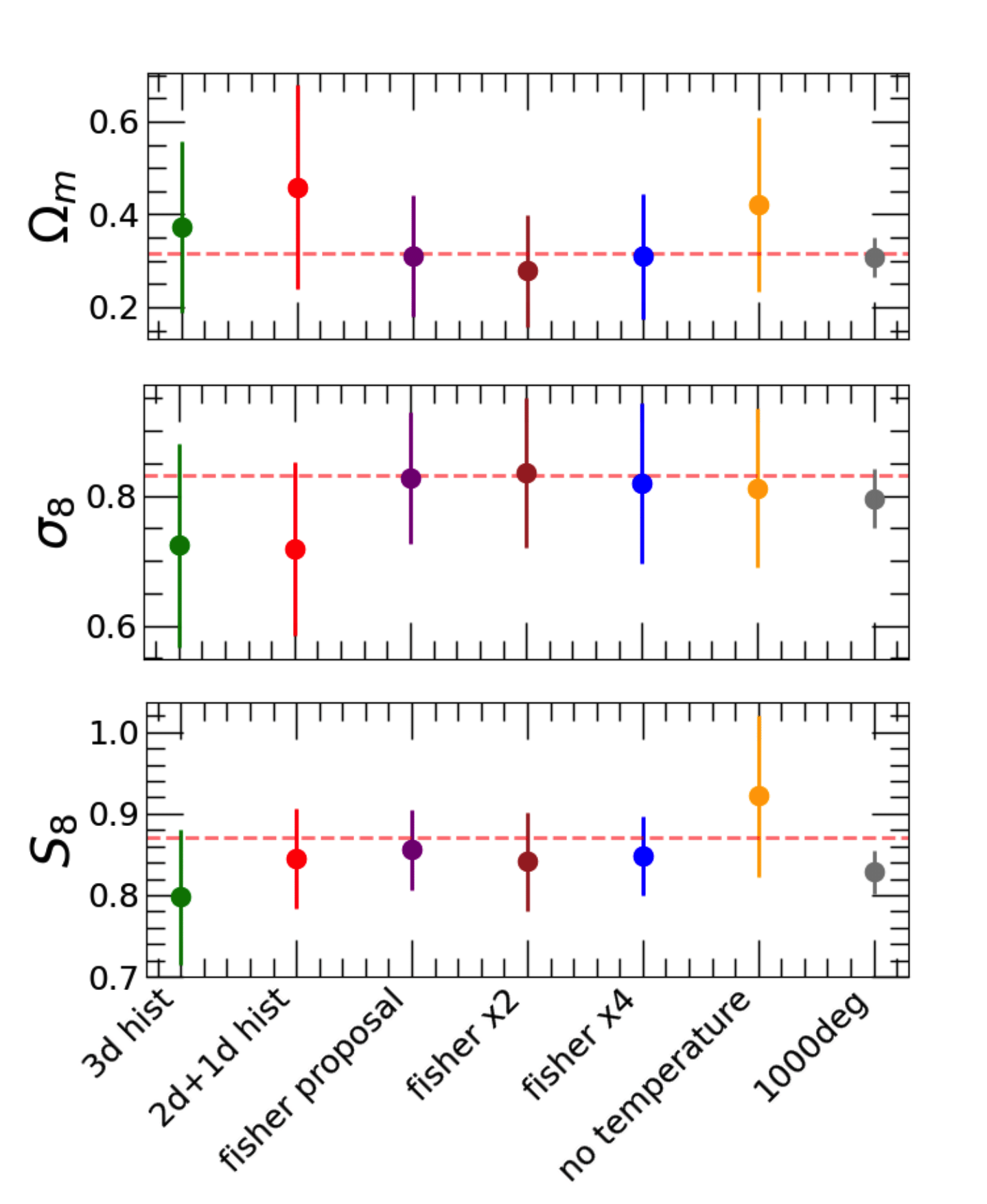}
\caption{Comparison of different SBI configurations in terms of $\Omega_m$, $\sigma_8$, and $S_8$. The \textit{3D hist} configuration outputs a normalized 3D histogram of temperature, flux, and redshift. The \textit{2D+1D hist} configuration provides a normalized 2D histogram of temperature and flux, along with a separate redshift histogram. The \textit{fisher proposal} configuration uses a non-uniform Fisher proposal for training, coming from a run on a mock sample covering 1000 deg$^2$. The \textit{fisher x2} and \textit{fisher x4} configurations apply non-uniform Fisher proposals for training, with factors of 2 and 4, respectively. The \textit{no temperature} configuration only includes flux and redshift in its output. Finally, the \textit{1000 deg$^2$} configuration uses a mock sample covering 1000 deg$^2$, whereas all other tests use a mock sample of 50 deg$^2$}
\label{Fig:sbi_confs}
\end{figure}

Figure \ref{Fig:sbi_confs} compares seven configurations in terms of the two cosmological parameters $\Omega_m$ and $\sigma_8$, as well as $S_8 \equiv \sigma_8 \cdot (\Omega_m / 0.27)^{0.3}$. SBI requires fixed-size outputs, but the number of generated clusters varies, which is why we chose to use fixed-bin histograms of temperature, flux, and redshift as outputs. We tested two different configurations to account for the temperature-flux correlation: a flattened 3D histogram, and a normalized 2D histogram of temperature and flux along with a separate redshift histogram, which ensures that the number of objects is counted only once. The 3D histogram contains more information, but most of the cells turn out to be empty, which renders optimization more challenging. Conversely, the 2D histogram solution loses information on the redshift evolution of the luminosity-temperature relation, but the optimization algorithm converges faster. Our tests show that both configurations converge to parameters that accurately reproduce the observed distributions of temperature, flux, and redshift. We then implemented a multivariate normal distribution as the proposal for training the model, see Appendix \ref{sec:Fisher proposal}. With this non-uniform proposal, the fit improves and convergence is faster, see Fig.\ref{Fig:proposal_test_2}. Since the statistical error and cosmic variance are smaller for 1000 deg$^2$ compared to 50 deg$^2$, we apply a scaling factor of 20 to the variance of each parameter from the 1000 square degree model to make it consistent with a 50 square degree model. We evaluate the impact of this scaling factor by applying values of 1, 2, and 4 to the Fisher matrix. All three Fisher matrix configurations converge, and the scaling factor does not seem to have a significant impact on the results. As discussed in Appendix \ref{sec:Impact of temperature}, we assess the effect of including the temperature distribution by running a test that only considers the flux and redshift distributions. Once again, we observe a significant difference in the error bars, particularly for the $S_8$ parameter, highlighting the importance of including temperature distributions in our simulated sample. Finally, we ran another test over a larger area to observe the effect of a higher number of clusters. As expected, this results in improved precision, confirming that a larger sample would significantly reduce the uncertainties.

\section{eRASS1-like area comparison} \label{sec:eRASS1}

\begin{figure}[H]
\centering
\includegraphics[width=\columnwidth]{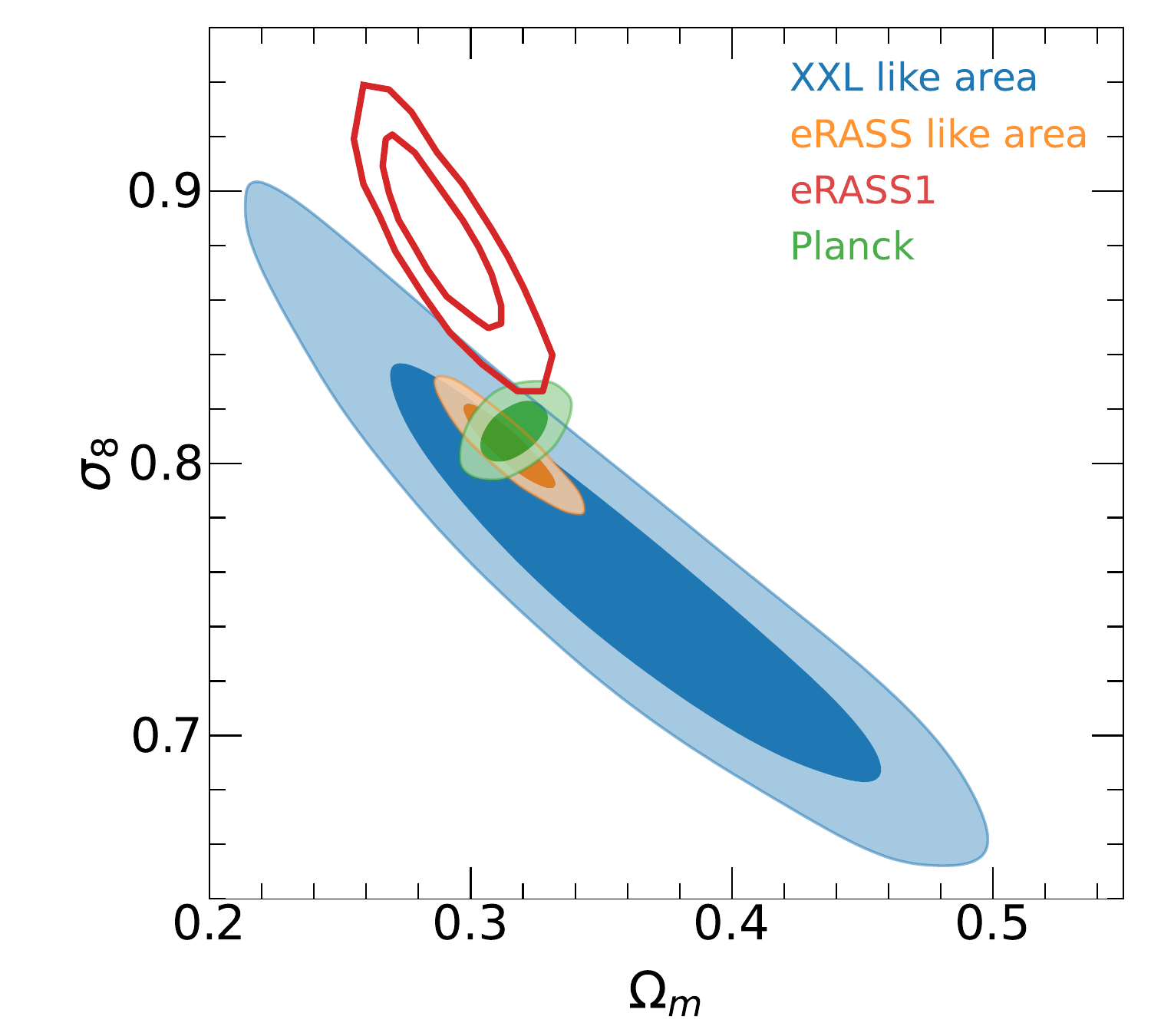}
\caption{Crossed posterior distributions of $\Omega_0$ and $\sigma_8$ for different survey areas. The distribution in \textit{blue} corresponds to a 50 deg$^2$ mock survey, while the \textit{orange} distribution represents a mock survey with a number of clusters comparable to eRASS1. The \textit{red} distribution shows the eRASS1 results from \citet{Ghirardini2024}, and the \textit{green} distribution corresponds to the Planck values.}
\label{Fig:eRASS1}
\end{figure}

Figure \ref{Fig:eRASS1} shows the simulation of a mock area with a number of clusters comparable to what is expected from eROSITA, in order to demonstrate a potential application of this method to the eRASS1 survey. Although eRASS1 covers a much larger area (8000 deg$^2$), it is approximately 10 times shallower than XMM-XXL; for this reason, we base the comparison on the number of detected clusters rather than on the survey area.
We compared the results with those from eRASS1 in \citep{Ghirardini2024}, as well as with results from this work on an XXL-like area and the Planck values used in the mock sample. The results show very low uncertainties and are quite promising. However, this assumes that we can constrain the temperature distribution of the survey, which we have not yet achieved. Without this temperature distribution, the uncertainties would be significantly larger as shown in Appendix \ref{sec:Impact of temperature}. Despite not having strong temperature constraints in eROSITA, we do have some insight into the temperature distribution because we have data from multiple energy bands. 

\end{document}